\documentclass{emulateapj}

\usepackage{graphicx}
\usepackage{subfigure}
\usepackage{verbatim}
\usepackage{natbib}
\newcommand{\htwo}{H$_2$}
\newcommand{\HI}{H{\footnotesize{I}}}
\newcommand{\HII}{H{\footnotesize{II}}}

\newcommand{\acet}{C$_2$H$_2$}

\newcommand{\cotwo}{CO$_2$}

\newcommand{\neII}{[\ion{Ne}{2}]}
\newcommand{\neIII}{[\ion{Ne}{3}]}
\newcommand{\sIII}{[\ion{S}{3}]}
\newcommand{\sIV}{[\ion{S}{4}]}
\newcommand{\arII}{[\ion{Ar}{2}]}
\newcommand{\arIII}{[\ion{Ar}{3}]}

\newcommand{\neV}{[\ion{Ne}{5}]}
\newcommand{\feII}{[\ion{Fe}{2}]}
\newcommand{\oIV}{[\ion{O}{4}]}
\newcommand{\siII}{[\ion{Si}{2}]}

\newcommand{\nII}{[\ion{N}{2}]}
\newcommand{\um}{$\mu$m}
\newcommand{\amir}{$\alpha_{30-20}$}
\newcommand{\anir}{$\alpha_{15-6}$}

\shorttitle{MID-IR SPECTROSCOPY OF COMPACT SYMMETRIC OBJECTS}
\shortauthors{WILLETT ET AL.}

\begin{document}

\title{Spitzer mid-IR spectroscopy of compact symmetric objects: What powers radio-loud AGN?}

\author{Kyle W. Willett\altaffilmark{1}, John T. Stocke\altaffilmark{1}, Jeremy Darling\altaffilmark{1}, \& Eric S. Perlman\altaffilmark{2}}

\altaffiltext{1}{Center for Astrophysics and Space Astronomy, Department of Astrophysical and Planetary Sciences, UCB 391, University of Colorado, Boulder, CO 80309-0391}
\altaffiltext{2}{Florida Institute of Technology, Physics and Space Sciences Department, 150 West University Boulevard, Melbourne, FL 32901}


\begin{abstract}
We present low- and high-resolution mid-infrared spectra and photometry for eight compact symmetric objects (CSOs) taken with the Infrared Spectrograph on the \textit{Spitzer Space Telescope}. The hosts of these young, powerful radio galaxies show significant diversity in their mid-IR spectra. This includes multiple atomic fine-structure lines, \htwo~gas, PAH emission, warm dust from T~=~50 to 150~K, and silicate features in both emission and absorption. There is no evidence in the mid-IR of a single template for CSO hosts, but 5/8 galaxies show similar moderate levels of star formation ($<10M_\sun$~yr$^{-1}$ from PAH emission) and silicate dust in a clumpy torus. The total amount of extinction ranges from $A_V\sim$~10~to~30, and the high-ionization \neV~14.3 and 24.3~\um~transitions are not detected for any galaxy in the sample. Almost all CSOs show contributions both from star formation and active galactic nuclei (AGNs), suggesting that they occupy a continuum between pure starbursts and AGNs. This is consistent with the hypothesis that radio galaxies are created following a galactic merger; the timing of the radio activity onset means that contributions to the infrared luminosity from both merger-induced star formation and the central AGN are likely. Bondi accretion is capable of powering the radio jets for almost all CSOs in the sample; the lack of \neV~emission suggests an advection-dominated accretion flow (ADAF) mode as a possible candidate. Merging black holes (BHs) with $M_{BH}>10^8M_\sun$ likely exist in all of the CSOs in the sample; however, there is no direct evidence from this data that BH spin energy is being tapped as an alternative mode for powering the radio jets. 
\end{abstract}




\section{Introduction}\label{sec-intro}

An enduring mystery in the study of radio galaxies is the nature of their power source. Specifically, most galaxies which possess a radio-luminous active galactic nucleus (AGN) and rapidly advancing jets show little evidence for a central accreting disk. For example, in Centaurus A and Virgo A (M87) there is no blue continuum or broad emission lines, but also no IR-emitting screen large enough to obscure a direct view of a typical broad emission line region \citep{per01a,why04,rad08}. While this modern paradigm unifies all AGN as supermassive black holes (BH) fed by an accretion disk orbiting the BH, many AGN show little evidence for the large amounts of circumnuclear gas needed to power them. These and other considerations led to the proposal that some AGN could be dominated by ``advection-dominated accretion flows'' \citep[ADAFs; e.g.][]{nar95}.

An alternative is that the active phase may be triggered by a major merger of disk galaxies and, concurrently, a merger of their supermassive BHs. It is now understood on the basis of detailed numerical simulations that the merger of disk galaxies can create a relatively normal elliptical galaxy \citep{bar92}. Since most (if not all) bright disk galaxies contain supermassive BHs, \citet{wil95} proposed a scenario for the creation of a radio-loud AGN in which the merger of two disk galaxies and their BHs can spin up the resulting BH, thus powering the AGN and its jets. The detection of X-ray point sources in radio-loud ellipticals provides observational evidence for the merging BH scenario \citep{kom03,bia08}. BH spin accounts naturally for the diminished output of successive radio outbursts in the brightest cluster galaxy (BCG) because the spin energy is diminished by each outburst and the accretion rate is too low for replenishment. At high redshift a merger of disk galaxies (a so-called ``wet merger'') and their individual supermassive BHs creates a rapidly-spinning supermassive BH and a luminous quasar.  As the cluster or group of galaxies in which the BCG is imbedded develops a hot, dense ICM, a large fraction of cluster galaxies are stripped of their cold and warm gas. As a result, all such mergers are ``dry'' and provide no additional accretion power. The BH spin hypothesis thus predicts the correct sign of radio-loud AGN evolution, which is not explained by the simplest accretion scenarios. 

An ideal class of galaxies in which to test these competing hypotheses are compact symmetric objects (CSOs). CSOs are radio-loud galaxies characterized by jet and/or hot spot activity on both sides of a central engine. They are morphologically similar to classical double radio sources, but with size scales of $<1$~kpc \citep{phi82,wil94}. The radio emission typically peaks near frequencies of a few GHz, which can be explained in most CSOs by synchrotron self-absorption in the galaxies \citep{dev09a,dev09}. Kinematic ages of the radio jets show they are extremely young, with ages of $\leq4000$~yr \citep{ows98,gug05}. As a result, it has been suggested that CSOs represent the early stages of an AGN evolving into an FR~II radio galaxy \citep{rea96a}, akin to a ``mini-Cygnus A'' \citep{beg96}. Given the young ages of these galaxies, they offer a much better chance for detecting signs of a recent merger that may be powering the radio activity. 

HST imaging of three nearby CSO host galaxies \citep{per01} show them to be nearly normal ellipticals; however, disturbed isophotes at the outer edges suggest that a past merger has taken place in all objects \citep{per01}. Simulations suggest that the radius of the photometric irregularities determines the timescale since the merger due to virialization of the stellar distribution \citep{mih95,mih96}. This is evidence for a recent merger in all three CSOs which commenced $\sim100$~million years ago, the time delay required for supermassive BHs to merge \citep{beg84}. This is also comparable to the timescale for driving gas to the center of the merger suggested by other simulations \citep{dim05,dim08,hop05,spr05,spr05a}. While this timescale supports the Wilson-Colbert hypothesis, it offers no concrete proof that radio-loud AGN are powered by BH spin. The accretion model requires large amounts of cold or warm gas close to the CSO, however, which is not seen - therefore, the power source for the jets is still an open question. 

Since distinguishing between a model for radio-loud AGN fueled by BH spin or one fueled by accretion will significantly affect our understanding of jet production and the differences between radio-loud and radio-quiet AGN, CSOs provide a unique opportunity to investigate the nuclear conditions in radio galaxies recently triggered by some mechanism into activity. One key question is whether large amounts of nuclear obscuration unresolved in the HST optical images could hide the accreting gas. Mid-IR observations should detect this gas if it exists, which may provide strong supporting evidence for accretion flows leading to AGN outbursts in CSOs. Observations like the HST imaging and the mid-IR results presented in this paper are the first that address such a hypothesis.

To this end we observed a sample of nearby CSOs with the {\it Spitzer Space Telescope} with the goal of distinguishing between these two hypotheses and characterizing the properties of CSO host galaxies, including gas and dust content and their star formation rates. If accretion is the power source for these galaxies, then mid-IR photons should penetrate any extinction screen that might obscure the AGN at shorter wavelengths; accreting gas should also emit via high-ionization atomic transitions due to the proximity of the AGN.  Absence of this would suggest a ``naked'' AGN with no luminous accreting gas, supporting a model in which radio-loud AGN are powered by BH spin due to either rapidly spinning BHs or binary BHs created in a merger. 

\begin{deluxetable*}{lcclcrclccrc}
\tabletypesize{\scriptsize}
\tablecaption{Properties of CSOs observed with Spitzer IRS \label{tbl-csosum}}
\tablewidth{0pt}
\tablehead{
\colhead{Object} & 
\colhead{RA} &
\colhead{Dec} & 
\colhead{z$_\sun$} &
\colhead{D$_L$} &
\colhead{H{\scriptsize{I}} column} &
\colhead{H{\scriptsize{I}}} & 
\colhead{Optical} &
\colhead{Optical} &
\colhead{log~P$_{radio}$} &
\colhead{log~L$_{IR}$} &
\colhead{log~L$_{X}$}
\\
\colhead{} & 
\colhead{J2000.0} & 
\colhead{J2000.0} & 
\colhead{} & 
\colhead{[Mpc]} &
\colhead{[cm$^{-2}$]} &
\colhead{ref.} & 
\colhead{spec.} & 
\colhead{ref.} & 
\colhead{[W/Hz]} &
\colhead{[L$_\odot$]} &
\colhead{[L$_\odot$]} 
}
\startdata
4C~+31.04                   &  01 19 35.0 & +32 10 50   & 0.0602 & 264  & $1.08\times10^{21}$    & (1)    & WLRG   & (8)    & 25.34 & 10.60  & \ldots \\
4C~+37.11                   &  04 05 49.2 & +38 03 32   & 0.055  & 242  & $1.8\times10^{20}$     & (2)    & NLRG   & (9)    & 25.05 & \ldots & \ldots \\
1146+59                     &  11 48 50.3 & +59 24 56   & 0.0108 & 48.2 & $1.82\times10^{21}$    & (1)    & LINER  & (10)   & 23.10 & 9.09   & \ldots \\
1245+676\tablenotemark{a}   &  12 47 33.3 & +67 23 16   & 0.1073 & 495  & $6.73\times10^{20}$    & (3)    & WLRG   & (8)    & 25.02 & \ldots & \ldots \\
4C~+12.50                   &  13 47 33.3 & +12 17 24   & 0.1217 & 571  & $T_s 6.2\times10^{18}$ & (4)    & NLRG   & (11)   & 26.30 & 12.15  & 43.3   \\
OQ~208                      &  14 07 00.4 & +28 27 15   & 0.0766 & 349  & $1.83\times10^{20}$    & (5)    & BLRG   & (12)   & 25.08 & 11.33  & 42.7   \\
PKS~1413+135                &  14 15 58.8 & +13 20 24   & 0.2467 & 1244 & $4.6\times10^{22}$     & (6)    & BL Lac & (13)   & 26.31 & 11.97  & 44.4   \\
PKS~1718-649                &  17 23 41.0 & $-$65 00 37 & 0.0142 & 62.6 & \ldots                 & \ldots & LINER  & (14)   & 24.25 & \ldots & \ldots \\
1946+70                     &  19 45 53.5 & +70 55 49   & 0.1008 & 460  & $2.2\times10^{23}$     & (7)    & \ldots & (15)   & 25.39 & \ldots & \ldots \\
\enddata
\tablenotetext{a}{Undetected with the spectral modules on the IRS; see \S\ref{ssec-no1245}.}
\tablerefs{
(1) - \citet{van89}; 
(2) - \citet{man04}; 
(3) - \citet{sai07};
(4) - \citet{mir89}; 
(5) - \citet{ver03a}; 
(6) - \citet{per02}; 
(7) - \citet{pec99}; 
(8) - \citet{mar96}; 
(9) - \citet{rod06};
(10) - \citet{kim89}; 
(11) - \citet{baa98};
(12) - \citet{pet07};
(13) - \citet{sto92};
(14) - \citet{fil85};
(15) - \citet{sne99}}
\end{deluxetable*}

In \S\ref{sec-obs}, we describe the CSO sample and the observations taken with {\it Spitzer}; we describe the data reduction process in \S\ref{sec-reduction}. \S\ref{sec-results} presents results of the observations, including full spectra and data tables for all measured features. We derive physical properties for the CSO hosts in \S\ref{sec-diagnostics} and classify them in the mid-IR. In \S\ref{sec-discussion} we discuss the evidence for the power source of CSOs and their role in the evolution of radio galaxies. An Appendix summarizes the jet properties and multi-wavelength observations of the host galaxies from the literature. We assume the WMAP5 cosmology with $H_0=70.5$~km~s$^{-1}$~Mpc$^{-1}$, $\Omega_M=0.27$, and $\Omega_\Lambda=0.73$. 



\section{CSO sample}\label{sec-obs}

Our goal for the sample was to select all known CSOs with $z<0.1$. This was motivated by the need for the galaxies to be IR-bright (detectable in reasonable integration times with the IRS) and having large amounts of multi-wavelength data to characterize the hosts. Radio surveys have detected CSOs out to redshifts of $z\sim0.5$, typically confirming their CSO status through VLBI \citep{pec00,tay03}.  In addition to objects selected at $z<0.1$ \citep{ode98,pih03,pol03}, we added the unusually IR-bright CSO PKS~1413+135, located at a redshift of $z=0.2467$. Recent IDs of CSOs at $z<0.1$ mean that the sample is likely not complete \citep[e.g.,][]{aug06}. 

The galaxies were observed in two separate programs during {\it Spitzer} cycles 3 and 5. The CSO 4C~+12.50 was previously observed with the IRS \citep{far07} and we used data publicly available through the {\it Spitzer} archive for this object.  The original sample also included IRS observations of the Seyfert 2 galaxy NGC~5793. Although the galaxy does possess parsec-scale symmetric radio emission \citep{hag00}, the radio spectrum is not gigahertz-peaked and it has no proper motion measurements that constrain the age. Data for this galaxy is thus not included in this paper. 

Table~\ref{tbl-csosum} lists the basic properties of the nine CSOs in the final sample. The heliocentric redshift is from optical measurements with the exception of PKS~1413+135, which uses the redshift from \HI~absorption \citep{car92}. Measurements of the \HI~column depend on the assumed spin temperature $T_s$, typically on the order of $\sim100$~K in these galaxies. Optical classifications of the hosts are: weak-line radio galaxy (WLRG), narrow-line radio galaxy (NLRG), low-ionization nuclear emission line region (LINER), BL Lac, or broad-line radio galaxy (BLRG). The radio luminosity is the specific power at 1.4~GHz from the NVSS \citep{con98}. The infrared luminosity is from IRAS fluxes and the prescription of \citet{san96}. The X-ray luminosity is measured in the 2--10~keV band (see Appendix for references).

\begin{deluxetable*}{llrrrrrrl}
\tabletypesize{\scriptsize}
\tablecaption{IRS observation log\label{tbl-obs}}
\tablewidth{0pt}
\tablehead{
\colhead{Object} & 
\colhead{Date} & 
\colhead{SL1} & 
\colhead{SL2} & 
\colhead{LL1} & 
\colhead{LL2} & 
\colhead{SH} & 
\colhead{LH} &
\colhead{Program}
}
\startdata
4C~+31.04    & 2007 Sep 01 & $60 \times 3$ & $60 \times 3$ & $120 \times  5$  & $120 \times 5$  & $120 \times 6$ & 240 $\times$ 4  & 30515 \\
4C~+37.11    & 2009 Apr 10 & $60 \times 3$ & $60 \times 3$ & $120 \times  5$  & $120 \times 5$  & $120 \times 9$ & 240 $\times$ 11 & 50591 \\
1146+59      & 2007 May 01 & $60 \times 2$ & $60 \times 2$ & $30  \times  3$  & $30  \times 3$  & $30  \times 6$ & 60  $\times$ 2  & 30515 \\
1245+676     & 2009 Apr 19 & $60 \times 4$ & $60 \times 4$ & $120 \times  6$  & $120 \times 6$  & $120 \times 5$ & 240 $\times$ 7  & 50591 \\
4C~+12.50    & 2004 Jan 07 & $14 \times 3$ & $14 \times 3$ & $30  \times  2$  & $30  \times 2$  & $30  \times 6$ & 60  $\times$ 4  & 105   \\
OQ~208       & 2007 Jun 24 & $60 \times 2$ & $60 \times 2$ & $30  \times  3$  & $30  \times 3$  & $30  \times 6$ & 60  $\times$ 3  & 30515 \\
PKS~1413+135 & 2007 Jul 31 & $60 \times 2$ & $60 \times 2$ & $30  \times  4$  & $30  \times 4$  & $120 \times 5$ & 60  $\times$ 6  & 30515 \\
PKS~1718-649 & 2009 Apr 18 & $14 \times 3$ & $14 \times 3$ & $30  \times  5$  & $30  \times 5$  & $30  \times 5$ & 60  $\times$ 5  & 50591 \\
1946+70      & 2009 Apr 07 & $60 \times 3$ & $60 \times 3$ & $120 \times  5$  & $120 \times 5$  & $120 \times 9$ & 240 $\times$ 12 & 50591 \\
\enddata
\tablecomments{Exposure times for all modules are given as seconds per cycle $\times$ number of cycles.}
\end{deluxetable*}


\section{{\it Spitzer} data reduction}\label{sec-reduction}

The IRS contains six different modules in both low- (LR) and high-resolution (HR) modes \citep{hou04}. The short-high (SH) and long-high (LH) modules operate at a resolution of $R\sim600$ with wavelength coverage from $9.9-37.2$~\um; the short-low (SL1 \& SL2) and long-low (LL1 \& LL2) modules operate at resolutions $R\sim56-127$, with coverage spanning $5.2-38$~\um. There also exist two ``bonus segments'' that cover the overlapping range between the 1$^{\textrm{\scriptsize{st}}}$ and 2$^{\textrm{\scriptsize{nd}}}$ orders in both SL and LL; we used these to verify the flux calibration between the first and second orders. 

In addition to the spectroscopic modules, the IRS also contains two peakup imaging arrays used to position the target on the spectral slit. We took dedicated 16 and 22~\um~sample-up-the-ramp (SUR) peakups for each object, giving two photometric data points used to calibrate the absolute flux scale of the spectra. The dedicated peakup observations were not coincident in time with the spectroscopy. Half of our sources are unresolved in the IRS slit and can be treated as point sources; for the remaining objects, we centered the slit along the brightest nucleus identified in the optical/near-IR. 

All objects in the sample were observed in the Staring Mode Astronomical Observing Template (AOT) with the galaxies placed at two nod positions located approximately one-third and two-thirds of the length of the slit. We were able to subtract out background sky contributions in all modules for most targets; the CSOs observed in 2007 have no sky observations for the SH module. Exposure times were chosen to yield signal-to-noise ratios of $S/N = 50$ for the low-resolution modules and $S/N\ge20$ for the continuum in the high-resolution modules (Table~\ref{tbl-obs}). 

The data were processed using the \textit{Spitzer} Science Center S17.0 data pipeline. We used the basic calibrated data products, already corrected for flat-fielding, stray light contributions, non-linear responsivity in the pixels, and ``drooping'' (increases in detector pixel voltage during non-destructive readouts). The 2-D images were first medianed at each nod position to remove transient effects such as cosmic rays. For the SL and LL modules, we subtracted the sky contribution by differencing the images for each nod with the adjacent position in the same module. 

Following the initial cleaning, further rogue pixels were eliminated using IRSCLEAN\_MASK\footnote{IRSCLEAN\_MASK is available at \\ \texttt{http://ssc.spitzer.caltech.edu/archanaly/\\contributed/irsclean/IRSCLEAN\_MASK.html}}. We used rogue pixel masks for each IRS campaign and supplemented the standard masks with manual cleaning of each nod and module. The 1-D spectra were then extracted using the Spitzer IRS Custom Extractor (SPICE) v.2.0. We used the optimal extraction routine with the standard aperture to improve the S/N ratio. 

The low-resolution modules were stitched together to match their continuum levels by using a multiplicative scaling. The flux level of the LL1 module was kept fixed, since this module has the most overlap with the transmission filter of the 22~\um~peakup. We then scaled the LL2 module to LL1, SL1 to the scaled LL2, and SL2 to the scaled SL1. The average modular scaling factors were $0.98$ for LL2 to LL1, $1.11$ for SL1 to LL2, and $1.16$ for SL2 to SL1. We then calibrated the low-resolution spectra as a single unit by matching the convolved flux to the IRS 22~\um~dedicated peakup using the method of \citet{arm07}. For 4C~+12.50, we scaled the IRS data to the IRAS 25~\um~flux (10\% uncertainty) with a scaling factor of 1.28. The required scaling in the majority of cases was quite small, indicating that the sky subtraction and spectral extraction techniques are robust; the mean overall scaling factor for the IRS peakups was $1.04$. Final spectra were produced from an error-weighted mean of the two nods. 

Noisy areas on both ends of the SH and LH orders were trimmed from the final spectra. These areas typically encompass a range of 10--30 pixels on the edge of the orders and correspond to areas of decreased sensitivity on the detector. We trimmed only pixels with an overlapping wavelength range in adjacent orders so that the maximum amount of information is preserved. 

\subsection{1245+676 non-detection}\label{ssec-no1245}

One of the CSOs, 1245+676, lies at a distance of 495~Mpc (the third-most distant galaxy in the sample) and was predicted to have a flux density of $\sim10$~mJy near 12~\um, based on a linear interpolation between existing optical/near-IR and radio photometry. This same method yielded flux estimates within a factor of a few for the other three CSOs in the sample without IRAS detections. The SUR peakup of 1245+676, however, only yielded flux densities of 0.7 and 1.0~mJy at 16 and 22~\um. The source of the flux estimation error likely comes from the 2MASS near-IR photometry, which may sample near the peak of an older population of red stars. If this is the case, then the true near-IR/optical continuum levels are significantly lower, which leads to a steeper optical-radio index and a lower estimate for the flux at 12~\um. 

The order of magnitude deficit in the expected flux means that the integration times for the IRS were too short, and the extracted spectra have no identifiable features (at 15~\um, the LR modules have S/N$<3$ and the SH module has S/N$<1$). The peakup measurements in Table~\ref{tbl-puflux} are regarded as valid, but all features in the 1-D spectra are limits (including the spectral indices, which are consistent with no slope). Therefore, we do not include data from this galaxy in subsequent analyses. 


\section{Results}\label{sec-results}

\begin{figure*}
\includegraphics[scale=1.0]{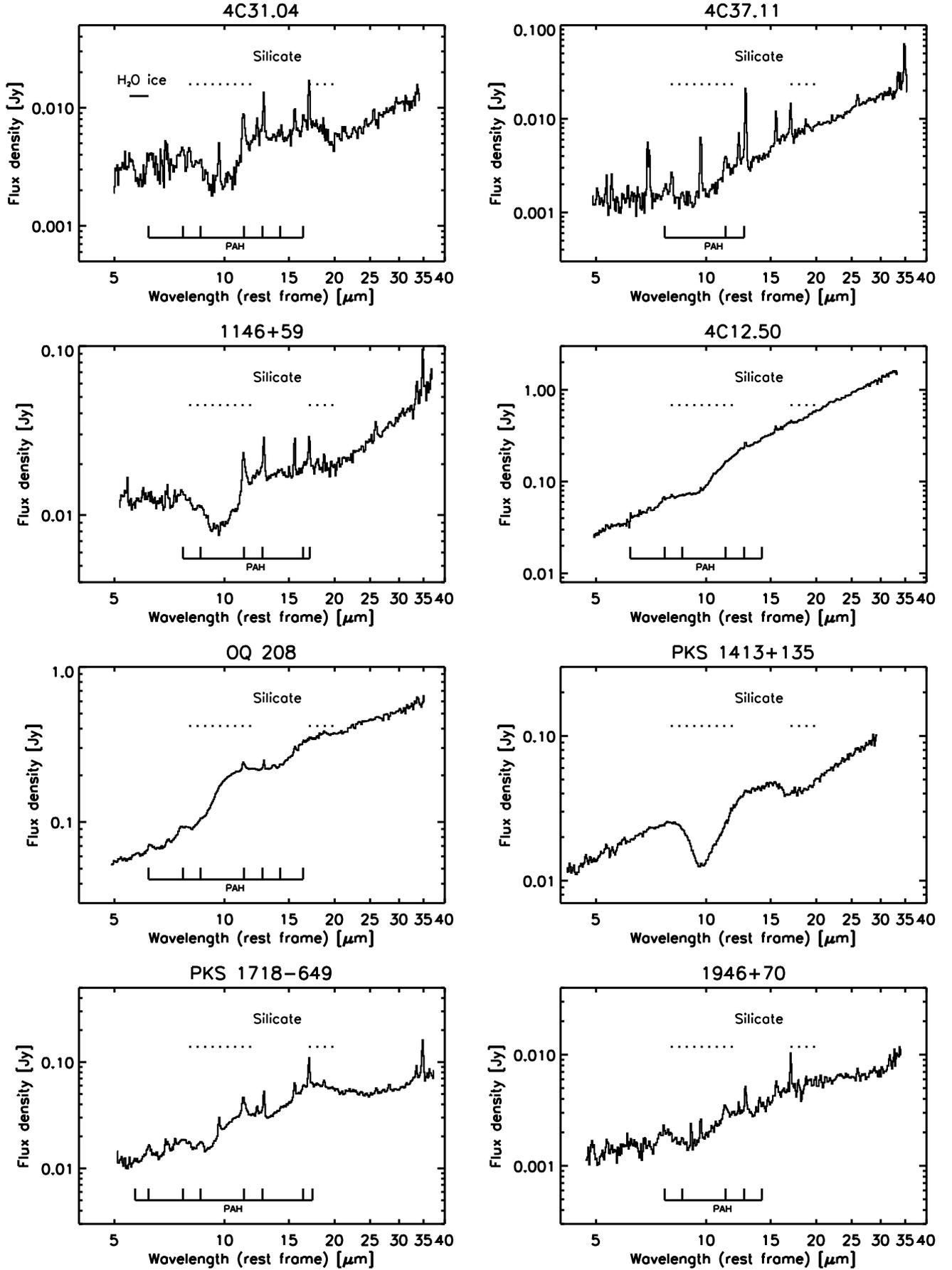}
\caption{Full low-resolution (LR) IRS spectra of the CSOs, with detections of PAHs, silicates and water ice features marked for each object.}\label{fig-lrspectra}
\end{figure*}

\begin{figure*}
\includegraphics[scale=1.0]{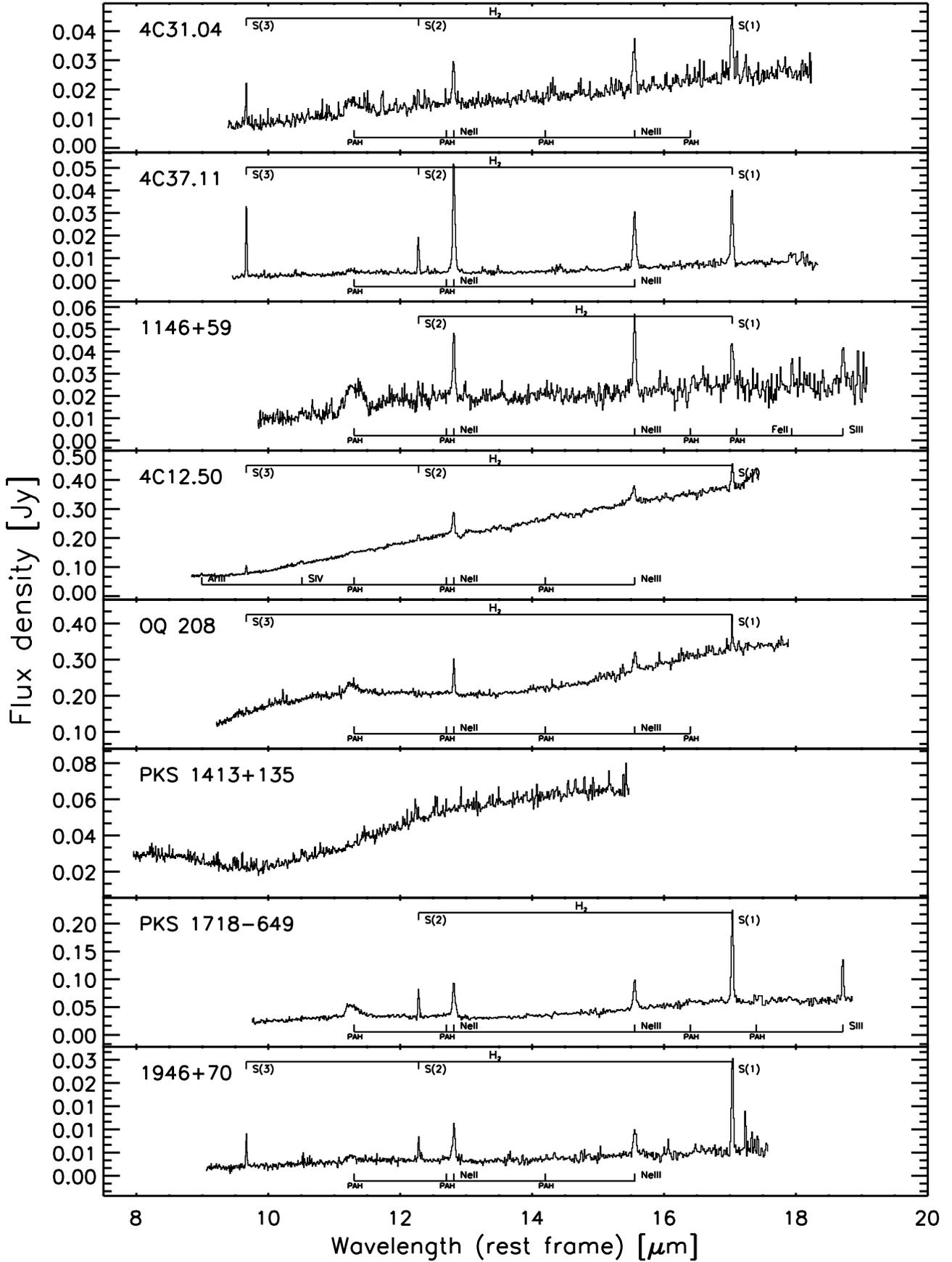}
\caption{Short-high (SH) spectra of the CSOs. Detected fine-structure atomic ({\it bottom}) and \htwo~({\it top}) emission lines are marked on each spectrum. PAH features are also marked along the bottom, although the broader profiles are more easily seen in the LR modules.}\label{fig-shspectra1}
\end{figure*}

\begin{figure*}
\includegraphics[scale=1.0]{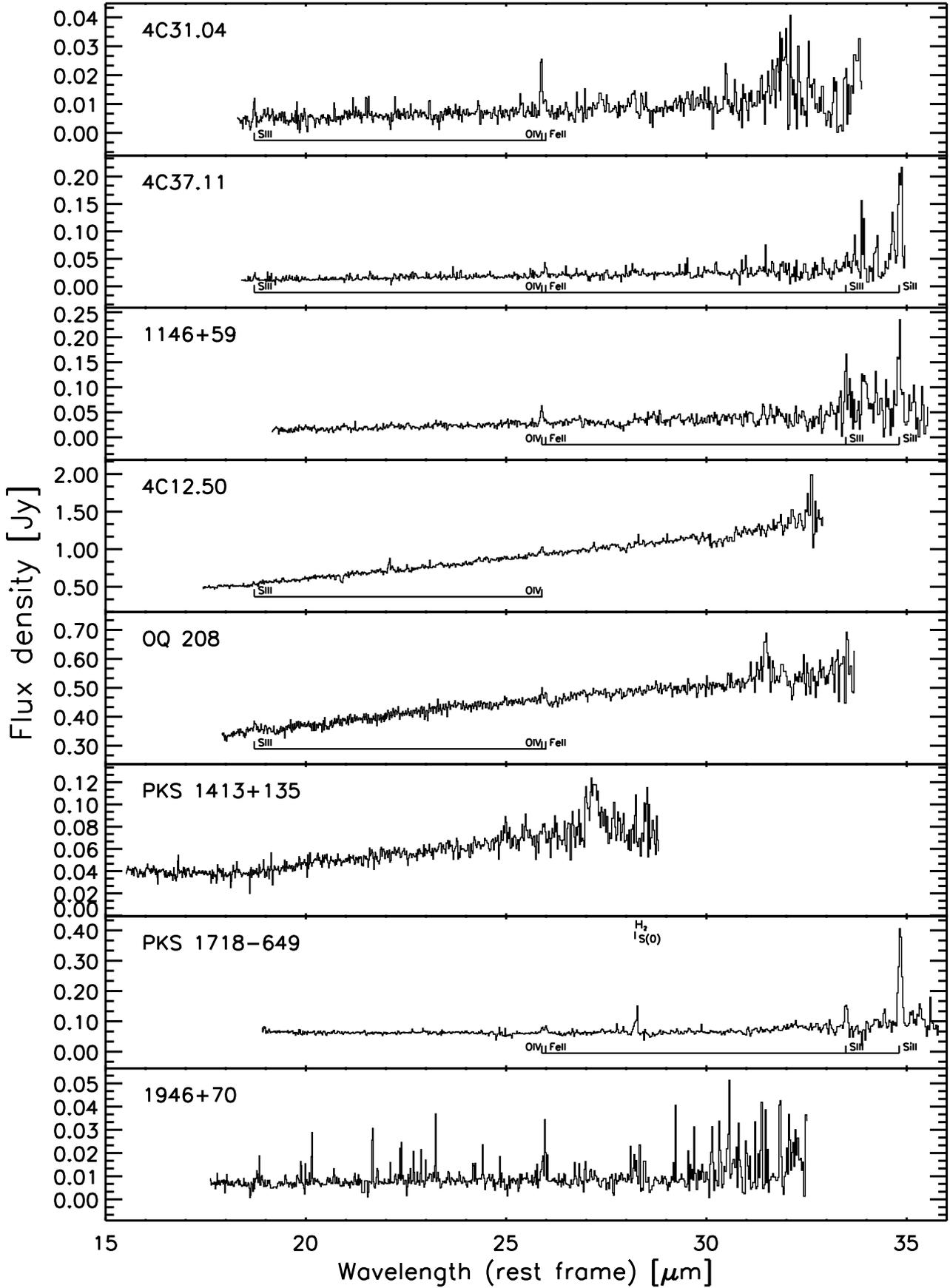}
\caption{Long-high (LH) spectra observed with the {\it Spitzer} IRS. Symbols are the same as Fig.~\ref{fig-shspectra1}. Features that are not labeled (such as the peak near $\lambda_{rest}=27$~\um~for PKS~1413+135) are considered spurious.}\label{fig-lhspectra1}
\end{figure*}

We show the photometrically-scaled low-resolution spectra in Figure~\ref{fig-lrspectra}, with individual modules stitched together. The SH spectra are shown in Figure~\ref{fig-shspectra1} and the LH spectra in Figure~\ref{fig-lhspectra1}. While individual orders within the high-resolution modules were typically well-aligned in flux, the differences in calibration between the SH and LH modules are clearly apparent when matching the spectra; this is due to a combination of different slit sizes for the SH and LH modules (by a factor of $\sim4$) and the lack of separate sky subtraction for the SH modules in some objects. For this reason, continuum levels between the SH and LH modules do not always match; the LH module is typically the more accurate measurement due to its removal of sky background in all galaxies. 

There still exists significant noise in the reduced spectra, most likely due to hot pixels or other instrumental artifacts from the IRS that were not corrected during the cleaning process. This occurs most often in the LH spectra - an example resembling a wide emission feature occurs between the 11$^{\textrm{\scriptsize{th}}}$ and 12$^{\textrm{\scriptsize{th}}}$ orders (see PKS~1413+135 near $\lambda_{rest}=27$~\um~in Figure~\ref{fig-lhspectra1}). We mark the locations of genuine features on the spectra themselves, with all detections we regard as real listed in Tables~\ref{tbl-pahsil}, \ref{tbl-atomic}, and \ref{tbl-h2}. All other features are considered spurious. 


\subsection{Continuum}\label{ssec-continuum}


\begin{figure}
\includegraphics[scale=0.5]{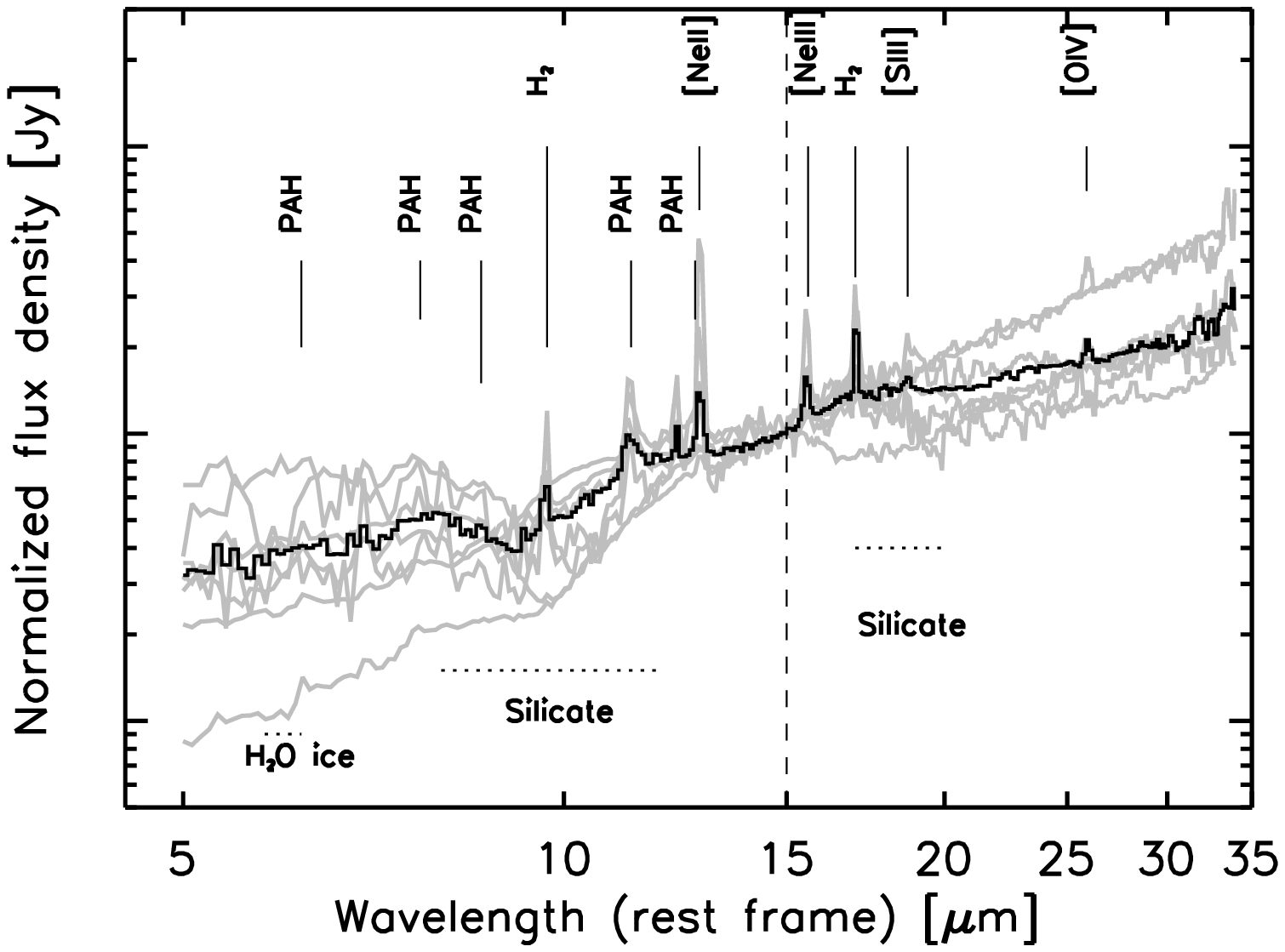}
\caption{Low-resolution spectra (grey) of the CSO sample, shifted to rest frame and normalized in flux at a line-free area at 15~\um. The thick spectrum is a composite template made from the error-weighted average of the individual scans, interpolated over a wavelength grid of $\Delta\lambda~=~0.08$~\um. \label{fig-lravg}}
\end{figure}

%
%

We measured the slope of the continuum in log space to create sets of rest-frame spectral indices for all objects (Table~\ref{tbl-puflux}). The continuum is characterized in two parts: a near-IR index between 6 and 15~\um~($\alpha_{15-6}$) and a mid-IR index between 20 and 30~\um~($\alpha_{30-20}$). All indices are measured in regions of line-free continuum, verified by visual examination for each object. The slope of the continuum is closely related to the shape of the underlying dust spectrum dominated by reprocessed thermal radiation from dust. IRS spectra from the larger sample of \citet{hao07} show a distinct change in spectral slopes according to their optical classifications: quasars have the shallowest \amir~and \anir, with AGN (Seyfert 1 and 2) slopes growing steeper and ULIRGs having the steepest slopes in both regimes. 

The CSOs have a wide range of spectral indices; \anir~ranges from 0.3 to 2.3, and \amir~from 0.0 to 2.0. The steepest \anir~and \amir~are both in 4C~+12.50, which has indices consistent with a single power law for the entire mid-IR spectrum. 1146+59 has the shallowest \anir~index and is flat between 5 and 8~\um. PKS~1718$-$649 has the shallowest \amir, also consistent with a flat spectrum. \citet{wu09} show that Seyfert 2 galaxies have steeper \amir~($1.53\pm0.84$) than in Seyfert 1 ($0.85\pm0.61$). The two Sy2 galaxies in the sample (4C~+37.11 and 4C~+12.50) have \amir~of 1.8 and 2.0, while the only Sy1 CSO in the sample (OQ 208) has \amir=0.9, fully consistent with this trend. The average indices for the sample are \anir$=1.1\pm0.6$ and \amir$=1.3\pm0.8$. 

Figure~\ref{fig-lravg} shows overlays of the LR spectra for all nine CSOs, with fluxes normalized in the feature-free region near 15~\um, along with a composite template medianed over each channel. The continuum emission of the individual galaxies varies most strongly longward of 15~\um; while most galaxies have \amir~consistent with the template spectral index of 1.0, 4C~+12.50 and 4C~+37.11 are both significantly above the average. The steepness of \anir~for 4C~+12.50 also strongly stands out, with the normalized spectrum falling below the template and all other galaxies shortward of 9~\um.


\begin{deluxetable}{lcccc}
\tabletypesize{\scriptsize}
\tablecaption{Peakup photometry and mid-IR spectral indices \label{tbl-puflux}}
\tablewidth{0pt}
\tablehead{
\colhead{Object} & 
\colhead{16 \um~PU} &
\colhead{22 \um~PU} & 
\colhead{$\alpha_{15-6}$} &
\colhead{$\alpha_{30-20}$} 
\\
\colhead{} & 
\colhead{[mJy]} & 
\colhead{[mJy]} & 
\colhead{} 
}
\startdata
4C~+31.04       & 4.8    & 6.3    & $0.6\pm0.3$ & $1.8\pm0.4$ \\
4C~+37.11       & 5.5    & 9.6    & $1.1\pm0.6$ & $1.8\pm0.3$ \\
1146+59         & 18     & 24     & $0.3\pm0.2$ & $1.7\pm0.3$ \\
1245+676        & 0.74   & 1.0    & --          & --          \\
4C~+12.50       & \ldots & \ldots & $2.3\pm0.2$ & $2.0\pm0.3$ \\
OQ~208 	        & 234    & 396    & $1.4\pm0.1$ & $0.9\pm0.1$ \\
PKS~1413+135    & 35     & 47     & $1.0\pm0.1$ & $1.8\pm0.4$ \\
PKS~1718-649    & 49     & 57     & $1.2\pm0.2$ & $0.0\pm0.1$ \\
1946+70         & 4.4    & 6.0    & $1.0\pm0.2$ & $0.4\pm0.4$ \\
\enddata
\tablecomments{Errors in peakup (PU) fluxes are at the 15\% level. 1245+676 has no spectral index measurements (see \S\ref{ssec-no1245}).} 
\end{deluxetable}

\subsection{PAH features}\label{ssec-pah}

Polycyclic aromatic hydrocarbons (PAHs) are among the dominant features in mid-IR spectra of dusty galaxies \citep{hel00,des07,smi07}; the emission is thought to be caused by UV excitation of vibrational modes in large molecules that are re-radiated in the infrared. We measured the strength of the primary PAH features using two different approaches: the first fits a spline to the local continuum \citep{spo07,zak08} and then integrates the total flux after a baseline has been subtracted. The second method uses PAHFIT \citep{smi07}, an IDL routine which performs a multi-component global fit on LR spectra based on a template for galaxies with a wide range of star formation. The advantage of PAHFIT is that it takes into account significant flux from the broad wings of dust features that blend with continuum in the spline-fit method; this is especially critical in measuring the equivalent widths (EWs) of PAH features. In order to compare the CSO hosts to other datasets \citep[e.g., the ``fork'' diagram in][]{spo07}, however, we follow the majority of the literature in using the spline-fit method. Results from both approaches are listed in Table~\ref{tbl-pahsil}. 


\begin{deluxetable*}{lrrrrrrrrrrrr}
\tabletypesize{\scriptsize}
\tablecaption{PAH and silicate features in low-resolution spectra\label{tbl-pahsil}}
\tablewidth{0pt}
\tablehead{
\colhead{} & 
\multicolumn{2}{c}{\underline{PAHFIT luminosity}} &
\multicolumn{3}{c}{\underline{PAHFIT EW}} &
\multicolumn{2}{c}{\underline{Spline-fit luminosity}} &
\multicolumn{3}{c}{\underline{Spline-fit EW}} &
\multicolumn{2}{c}{\underline{Silicate}}
\\
\colhead{Object} & 
\colhead{6.2} &
\colhead{11.3} & 
\colhead{6.2} &
\colhead{6.2 ice} &
\colhead{11.3} &
\colhead{6.2} &
\colhead{11.3} & 
\colhead{6.2} &
\colhead{6.2 ice} &
\colhead{11.3} &
\colhead{$S_{9.7}$} &
\colhead{$S_{18}$}
\\
\colhead{} & 
\colhead{[log $L/L_\sun$]} & 
\colhead{[log $L/L_\sun$]} & 
\colhead{[\um]} & 
\colhead{[\um]} & 
\colhead{[\um]} &
\colhead{[log $L/L_\sun$]} & 
\colhead{[log $L/L_\sun$]} & 
\colhead{[\um]} & 
\colhead{[\um]} & 
\colhead{[\um]} &
\colhead{} &
\colhead{}
}
\startdata
4C~+31.04    &  8.02    & 8.19    & 0.18    & 0.16    & 0.62    & 7.88      & 7.81      & 0.18      & 0.12    & 0.32      & $-0.8$    & $-0.4$  \\
4C~+37.11    &  $-$     & 7.56    & $-$     & $-$     & 0.28    & $<7.21$   & 7.25      & $<0.08$   & $-$     & 0.17      & $-0.5$    & $-0.0$  \\
1146+59	     &  6.61    & 7.01    & 0.05    & $-$     & 0.37    & 6.00      & 6.71      & 0.01      & $-$     & 0.23      & $-0.7$    & $-0.2$  \\
4C~+12.50    &  9.37    & 9.43    & 0.07    & $-$     & 0.05    & 8.20      & 8.73      & 0.01      & $-$     & 0.01      & $-0.5$    & $-0.0$  \\
OQ~208 	     &  9.24    & 9.36    & 0.08    & $-$     & 0.09    & 8.75      & 8.69      & 0.03      & $-$     & 0.03      & $+0.3$    & $+0.2$  \\
PKS~1413+135 &  $-$     & $<8.82$ & $-$     & $-$     & $<0.02$ & $<9.05$   & $<8.76$   & $<0.01$   & $-$     & $<0.01$   & $-1.0$    & $-0.3$  \\
PKS~1718-649 &  7.11    & 7.49    & 0.08    & $-$     & 0.30    & 6.67      & 7.08      & 0.04      & $-$     & 0.13      & $-0.4$    & $+0.3$  \\
1946+70      &  $-$     & 8.04    & $-$     & $-$     & 0.22    & $<7.48$   & 7.58      & $<0.04$   & $-$     & 0.09      & $-0.6$    & $-0.3$  \\
\enddata
\tablecomments{The 6.2~\um~PAH ``ice'' data use a continuum that has been corrected for water ice absorption (where present) at 6~\um. $-$ indicates that PAHFIT fit no significant flux for a particular dust component. Silicate strength is defined in Equation~\ref{eqn-silstrength}.}
\end{deluxetable*}

The strengths of the PAH features in the CSOs vary considerably, with no uniform template typical for the sample (Figure~\ref{fig-lravg}). PKS~1413+135 shows no evidence for PAHs in any of the five main bands commonly seen in ULIRGs (6.2, 7.7, 8.6, 11.3, and 12.7~\um). This may be an indication of the hard UV/X-ray environment that is known to dissociate PAHs \citep{voi92}, which can arise from either extreme star formation or direct exposure to a central power source such as an AGN. It could also signify a lack of gas/dust near the ionizing source. No other tracers of star formation (such as Ly-$\alpha$ or fine-structure lines in the optical/IR bands) are identified in PKS~1413+135, suggesting that either the radiation of the AGN or large amounts of obscuration are responsible for the lack of PAH detections in this object. 

4C~+12.50 and OQ~208 both show weak PAH emission above a strong continuum. The 11.3 and 12.7~\um~features are strongest in OQ~208, while the PAH emission in 4C~+12.50 is very dim for all detections. PAH features are seen in all five main bands for both objects, in addition to bands at 14.2 and 16.4~\um~in OQ~208. We note that PKS~1413+135, OQ~208, and 4C~+12.50 are all strong X-ray sources with $L_{2-10~keV}>10^{42}$~erg/s, further suggesting that dissociation of the molecules may be the primary reason for weak emission in these galaxies. 

1146+59, 4C~+31.04, 1946+70, PKS~1718$-$649, and 4C~+37.11 all show moderate PAH emission strongest at 11.3 and 12.7~\um. Two CSOs, 1946+70 and 4C~+37.11, have no detectable 6.2~\um~PAH emission, although the upper limit is consistent with the 6.2/11.3~\um~band ratio of 1.1 found by \citet{smi07} for star-forming galaxies. Previous studies of Galactic PAH bands show that the 6.2~\um~luminosity can be suppressed with respect to the other bands for multiple reasons, including differential extinction \citep{rig02} or hot dust increasing the continuum level and obscuring the wings of the PAH feature \citep{des07}. A neutral ISM may also affect their relative strengths, as the 6.2~\um~mode is stronger when the PAHs are ionized \citep{ver96}. 


\subsection{Silicate absorption}\label{ssec-sildust}

Another prominent feature in the LR spectra is the pair of amorphous silicate emission/absorption complexes centered at 9.7 and 18~\um. The strengths of the silicate features are measured using the method of \citet{spo07}:

\begin{equation}
\label{eqn-silstrength}
S_{\lambda} = ln\left(\frac{F_{\lambda}}{F_{cont}}\right),
\end{equation}

\noindent where $F_\lambda$ is the flux extremum near 9.7 or 18~\um~and $F_{cont}$ is the continuum flux interpolated at the same wavelength from a spline fit. Results for the silicate strength fall into two categories: emission and weak absorption (Table~\ref{tbl-pahsil}).

OQ~208 is the only object that shows silicate in emission, with $S_{9.7} = 0.3$ and $S_{18}=0.2$. The remainder of its mid-IR emission follows a relatively flat power law with an index of 1.9. \citet{hao07} found that silicate emission in galaxies occurred almost exclusively for quasars and Seyfert 1 galaxies, with an average $S_{9.7}=0.20$ for quasars and $S_{9.7}=-0.18$ for Seyfert 1 galaxies. The LR spectrum for OQ~208 closely resembles the average mid-IR quasar template generated from $\sim$20 objects in the \citet{hao07} sample; this suggests a relatively small amount of dust that is optically and geometrically thin \citep{lev07} to produce the emission feature. A clumpy medium has been shown to reproduce such features due to illuminated clouds behind the central source (relative to the observer) filling in the absorption trough sufficiently to create silicate emission \citep{nen02}. 

Six CSOs show weak absorption, all with silicate features ranging from $0.0<S_{9.7}<-0.8$. Narrow line emission from \htwo~S(3) and/or \sIV~can be seen on top of the silicate absorption in several objects, suggesting that at least some part of the narrow-line gas is foreground to the absorbing dust. The average silicate strength is consistent with that seen both in Seyfert galaxies and ULIRGs \citep{hao07}. Models from \citet{lev07} suggest that this type of emission may arise from dust clouds that are optically but {\em not} geometrically thick (ie, a clumpy ISM, characteristic of optically identified AGN). None of these galaxies show strong absorption from the secondary silicate feature at 18~\um. PKS~1413+135 shows the deepest silicate absorption at $S_{9.7}=-1.0$. No objects show the deep absorption features ($S_{9.7}<-1.5$) seen in Seyfert~2 galaxies and ULIRGs that can only be generated with a smooth, spherical shell of dust.  

The 9.7~\um~silicate depth is also translated into an apparent visual extinction using the Galactic calibration of \citet{roc84}; the measured extinction spans $A_V~\sim~10-30$~mags (Table~\ref{tbl-derived}).  


\subsection{Other absorption features: water ice, hydrocarbons, and gas-phase molecules}\label{ssec-abs}

\begin{deluxetable*}{lrrrrrrrrrrrr}
\tabletypesize{\scriptsize}
\tablecaption{Atomic line luminosities from HR spectra\label{tbl-atomic}}
\tablewidth{0pt}
\tablehead{
\colhead{Object} & 
\colhead{[Ar \tiny{III}\footnotesize]} &
\colhead{[S \tiny{IV}\footnotesize]} &
\colhead{[Ne \tiny{II}\footnotesize]} &
\colhead{[Ne \tiny{V}\footnotesize]} &
\colhead{[Ne \tiny{III}\footnotesize]} &
\colhead{[Fe \tiny{II}\footnotesize]} &
\colhead{[S \tiny{III}\footnotesize]} &
\colhead{[Ne \tiny{V}\footnotesize]} &
\colhead{[O \tiny{IV}\footnotesize]} &
\colhead{[Fe \tiny{II}\footnotesize]} &
\colhead{[S \tiny{III}\footnotesize]} &
\colhead{[Si \tiny{II}\footnotesize]}
\\
\colhead{$\lambda_{rest}$ [\um]} & 
\colhead{8.991} & 
\colhead{10.511} & 
\colhead{12.814} & 
\colhead{14.322} & 
\colhead{15.555} & 
\colhead{17.936} & 
\colhead{18.713} &
\colhead{24.318} &
\colhead{25.890} &
\colhead{25.988} &
\colhead{33.481} &
\colhead{34.815}
}
\startdata
4C~+31.04    &  $-$     & $<7.03$  & 7.34    & $< 6.89$ & 7.32    & $<6.53$ & 6.71    & $<6.72$ & 7.04    & 6.61    & $<7.39$ & $-$  \\
4C~+37.11    &  $-$ 	& $<6.71$  & 7.81    & $< 6.63$ & 7.43    & $<6.38$ & 6.97    & $<6.59$ & 6.43    & 6.93    & 7.30    & 7.93 \\
1146+59      &  $-$     & $<5.72$  & 6.19    & $< 5.51$ & 6.12    & 5.40    & 5.64    & $<5.24$ & 5.95    & 5.18    & 6.11    & 6.34 \\
4C~+12.50    &  7.76	& 8.19     & 8.66    & $< 8.01$ & 8.62    & $<8.02$ & 8.03    & $<8.33$ & 8.55    & $<8.55$ & $-$     & $-$  \\
OQ~208 	     &  $-$     & $<7.79$  & 8.22    & $< 7.66$ & 8.10    & $<7.39$ & 7.68    & $<7.54$ & 7.47    & 7.45    & $<8.33$ & $-$  \\
PKS~1413+135 &  $<8.34$ & $<8.58$  & $<8.44$ & $< 8.27$ & $<8.76$ & $<8.23$ & $<8.35$ & $<8.20$ & $<8.49$ & $<8.54$ & $-$     & $-$  \\
PKS~1718-649 &  $-$ 	& $<5.97$  & 6.80    & $< 5.66$ & 6.58    & $<5.47$ & 6.39    & $<5.55$ & 5.75    & 5.87    & 6.35    & 6.95 \\
1946+70      &  $<6.73$ & $<7.22$  & 7.58    & $< 6.74$ & 7.35    & $<6.94$ & $<7.32$ & $<7.41$ & $<7.56$ & $<7.62$ & $-$     & $-$  \\
\enddata
\tablecomments{All line luminosities are in log~$L_\sun$; $-$ indicates that the observed wavelength was not visible in the IRS range. Errors in the flux measurements are on the order of 10\%.}
\end{deluxetable*}

Water ice absorption at 6~\um~has been observed in the mid-IR in ULIRGs but is typically absent in Seyfert and starburst galaxies \citep{spo00,spo02}. We detect water ice in 4C~+31.04 at an optical depth of $\tau=0.57$. The feature is shallow and bracketed by possible PAH emission near 6.2~\um, which makes accurate measurements of the local continuum difficult. There is no indication of water ice in any of the other objects, nor in the stacked LR spectra of the CSOs (Figure~\ref{fig-lravg}); this argues against the existence of any weak ice features not seen in individual objects. 

Bending modes in hydrogenated amorphous carbon (HAC) \citep{spo01,dar07b,dar07c} cause two blended absorption features in the mid-IR centered at 6.85 and 7.25~\um. We detect both features at a weak level in 1146+59; we use a spline-fit to characterize the local continuum and measure their optical depths and integrated fluxes. The 6.85~\um~HAC has $\tau=0.12$ and $f=-0.7\times10^{-21}$~W~cm$^{-2}$, with the 7.25~\um~HAC at $\tau=0.10$ and $f=-0.6\times10^{-21}$~W~cm$^{-2}$. The CSO detection rate for HACs (1/8) is consistent with that seen in larger samples of IR-bright galaxies (Willett et~al. 2010), especially in those where strong emission from the 6.2~\um~PAH feature may obscure the HAC absorption.

Gas-phase molecules (including \acet, HCN, and \cotwo) have also been observed in absorption in galaxies in the mid-IR, typically in deeply obscured ULIRG nuclei \citep{lah07}. No detections of any of these molecules were made for the sample; given the characteristic column density needed to show absorption, however, the S/N is likely too low in all galaxies to be detected at a typical column density of $N\gtrsim10^{16}$~cm$^{-2}$.

\subsection{Atomic lines}\label{ssec-atomic}

The high-resolution data is primarily used for the identification of atomic and molecular emission lines with widths too narrow to be separated in the LR modules. Integrated luminosities (assuming isotropy) are given in Table~\ref{tbl-atomic}; upper limits are computed by assuming a Gaussian with a peak flux density $3\sigma$ above the noise and a velocity width derived from detections in the sample of ULIRGs from Willett et~al. (2010). Line fluxes were measured using the standard packages in the Spectroscopic Modeling Analysis and Reduction Tool (SMART) v6.2.4 \citep{hig04}. We found a simple Gaussian to be a good fit for virtually all high-resolution lines, with the central wavelength left as a free parameter in the fit; for cases in which lines are blended (such as the \oIV/\feII~complex), we used a two-component Gaussian fit centered at their expected wavelengths from optical redshifts. 


The fine structure \neII~12.8\um~and \neIII~15.6\um~transitions are the most commonly detected in the sample, with both observed in seven objects. We also detected \arIII~9.0\um, \sIII~18\um, \sIII~33\um, \sIV~10.5\um, \siII~35\um, \feII~23\um, \feII~26\um, and \oIV~25.9\um. The strongest lines seen in the HR modules, \neII~and \neIII, were resolved in 6/7 objects with neon detections. The velocity width for these transitions spans 250--600~km/s, with the broadest example the \neIII~transition in 4C~+12.50 (see \S\ref{ssec-bhmass} for a discussion of linewidths for this galaxy). The high-ionization \neV~line, commonly observed in AGN with accreting gas, was not detected in any object in either the 14.3 or the 24.3~\um~transition. 

As a check for our line measurements, we compared lines detected in the archival galaxy 4C~+12.50 with HR results published by \citet{far07}. For the most part, the two sets of measured fluxes are in good agreement; however, we cannot confirm detections of the \neV~14\um~or \htwo~S(0) transitions, both of which are considered uncertain by \citet{far07}. The authors also measure a flux for the \neIII~transition that is higher than our measurement by $\sim30\%$; this may be a result of different fitting techniques, since the line has a non-Gaussian component in a broad blue wing, possibly from a photoionized outflow \citep{spo09}. 



\subsection{Molecular hydrogen}\label{ssec-h2results} 


\begin{deluxetable*}{lrrrrrrrr}
\tabletypesize{\scriptsize}
\tablecaption{Line luminosities of molecular H$_2$ gas\label{tbl-h2}}
\tablewidth{0pt}
\tablehead{
\colhead{Object} & 
\colhead{H$_2$~S(7)} &
\colhead{H$_2$~S(6)} &
\colhead{H$_2$~S(5)} &
\colhead{H$_2$~S(4)} &
\colhead{H$_2$~S(3)} &
\colhead{H$_2$~S(2)} &
\colhead{H$_2$~S(1)} &
\colhead{H$_2$~S(0)}
\\
\colhead{$\lambda_{rest}$ [\um]} & 
\colhead{5.51 } & 
\colhead{6.11 } & 
\colhead{6.91 } & 
\colhead{8.03 } & 
\colhead{9.67 } & 
\colhead{12.28} & 
\colhead{17.04} & 
\colhead{28.22}
}
\startdata
4C~+31.04 	& $<7.34$ & $<7.25$  & 7.53:   & $<6.47$ & 7.37    & 6.87    & 7.46    & $<6.77$ \\
4C~+37.11     	&   7.07  &   6.66   & 7.70:   & 7.09    & 7.59    & 7.18    & 7.38    & $<6.98$  \\
1146+59 	& $<6.09$ & $<5.91$  & 6.08:   & $<5.23$ & $<5.29$ & 5.74    & 5.75    & $<5.49$ \\
4C~+12.50 	& $<8.53$ & $<8.38$  & $<8.84$ & $<7.88$ & 8.21    & 8.04    & 8.43    & $<8.55$  \\
OQ~208 		& $<8.07$ & $<7.93$  & 8.27:   & $<7.49$ & 7.31    & $<7.50$ & 7.86    & $<7.51$  \\
PKS~1413+135 	& $<8.80$ & $<8.62$  & $<8.79$ & $<8.24$ & $<8.38$ & $<8.41$ & $<8.22$ & $<8.79$  \\
PKS~1718-649   	& $<6.41$ & $<6.24$  & 6.69:   & $<5.57$ & $<5.83$ & 6.49    & 6.87    & 6.26    \\
1946+70       	& $<7.26$ & $<7.08$  & $<7.53$ & $<6.58$ & 7.44    & 7.16    & 7.67    & $<7.57$  \\
\enddata                       
\tablecomments{Luminosities are given in log~$L_\sun$. Measured H$_2$~S(5) lines (indicated with :) are upper limits due to possible blending with [Ar~\tiny{II}\scriptsize]~emission; see \S\ref{ssec-h2results} for details. The \htwo~S(4-7) lines were measured in the LR modules; all other lines were measured in the high-resolution modules. Errors in the flux measurements are on the order of 10\%.}
\end{deluxetable*}

Lines from the rotational series of \htwo~are seen in 7/8 CSOs. The exception, PKS~1413+135, showed no identified atomic or molecular emission lines at all in the mid-IR. The strongest line in all objects is the ortho \htwo~S(1)~17.0\um~transition, appearing in every object in which molecular hydrogen was detected. The next ortho transition, S(3)~9.67\um, was detected in five objects; we can only place upper limits on the S(5)~6.91\um~transition due to likely blending from \arII~6.99\um~in the low-resolution spectra. For the para \htwo~transitions, the ground state S(0)~28.2\um~was detected in only one object, although this line can be obscured by rising continuum levels near 28~\um. The next excited state, S(2)~12.3\um, is detected in six objects, with OQ~208 the only object in which the S(1) transition was observed without S(2). Fluxes for all lines are given in Table~\ref{tbl-h2}; upper limits are computed in the same manner as those for atomic emission lines. 

4C~+37.11 shows a highly unusual \htwo~spectrum, with all lines from S(1) through S(7) detected in the LR data. This is the only object that shows clear detections of the S(4)~8.03\um, S(6)~6.11\um, and S(7)~5.51\um~lines, and the \htwo~luminosities are much larger compared to the strength of its fine-structure lines (such as neon) than in any other object. In a sample of 77 ULIRGs analyzed by \citet{hig04}, the S(4)--S(7) lines were seen only in a single nearby target (NGC~6240). Similar \htwo~strengths are also observed in the FR~II radio galaxy 3C~326 \citep{ogl07a} and the compact group Stephan's Quintet \citep{app06}. Since other tracers such as PAH strength and \neII~emission are inconsistent with extreme amounts of star formation in this galaxy, the strong \htwo~lines in these objects are attributed to shock heating of the gas by a tidal accretion flow arising from interactions with other galaxies. \citet{man04} show that the radio structure in 4C~+37.11 may come from interactions driven by a recent merger, with a dense circumnuclear medium giving rise to shocks in the central component that would be consistent with the \htwo~lines seen here.


\section{Infrared diagnostics}\label{sec-diagnostics}


\begin{deluxetable*}{lrrcllccrcc}
\tabletypesize{\scriptsize}
\tablecaption{Derived mid-IR properties of the CSO sample\label{tbl-derived}}
\tablewidth{0pt}
\tablehead{
\colhead{Object} & 
\colhead{SFR$_{Ne}$} &
\colhead{SFR$_{PAH}$} &
\colhead{$A_V$} &
\colhead{$T_{H_2}^{warm}$} & 
\colhead{$M_{H_2}^{warm}$} &
\colhead{$T_{H_2}^{hot}$} & 
\colhead{$M_{H_2}^{hot}$} &
\colhead{T$_{dust}$} &
\colhead{log~$M_{BH}^{OIV}$} &
\colhead{log~$M_{BH}^{bulge}$}
\\
\colhead{} & 
\colhead{[M$_\sun$/yr]} &
\colhead{[M$_\sun$/yr]} &
\colhead{[mag]} &
\colhead{[K]} & 
\colhead{[$10^7 M_\sun$]} & 
\colhead{[K]} & 
\colhead{[$10^7 M_\sun$]} & 
\colhead{[K]} & 
\colhead{[$M_\sun$]} &
\colhead{[$M_\sun$]} 
}
\startdata
4C~+31.04 	 & $7.8\pm1.1$   & 6.4         & $15\pm6$   & $338\pm100$ & $0.47\pm0.13$  & $-$         & $-$           & 65    & $<8.16$       & 8.78 \\	
4C~+37.11     	 & $17\pm1$      & $[0.8-1.6]$ & $9\pm3$    & $354\pm90$  & $0.95\pm0.05$  & $854\pm240$ & $0.09\pm0.02$ & --    & $<7.96$       & 8.74 \\	
1146+59 	 & $0.5\pm0.1$   & 0.3         & $13\pm1$   & $537\pm50$  & $0.01\pm0.002$ & $-$         & $-$           & 150   & $9.23\pm0.52$ & 8.40 \\	
4C~+12.50 	 & $159\pm14$    & 31.5        & $8\pm0.6$  & $326\pm97$  & $4.26\pm0.46$  & $-$         & $-$           & 85    & \ldots        & 8.81 \\	
OQ~208 		 & $54\pm3$      & 47.8        & $-$        & $281\pm101$ & $1.13\pm0.13$  & $-$         & $-$           & 110   & $<7.79$       & 8.81 \\	
PKS~1413+135 	 & $<150$        & $<80$       & $13\pm0.4$ & $-$         &  $-$           & $-$         & $-$           & 90    & $-$           & 8.22 \\	
PKS~1718-649     & $1.8\pm0.1$   & 0.8         & $7\pm3$    & $202\pm112$ & $0.12\pm0.01$  & $-$         & $-$           & --    & $8.62\pm0.45$ & 8.48 \\	
1946+70       	 & $11.1\pm0.9$  & $[1.7-3.1]$ & $11\pm3$   & $320\pm83$  & $0.74\pm0.05$  & $-$         & $-$           & --    & $-$           & 8.54 \\	
\enddata
\tablecomments{Multiple values for $SFR_{PAH}$ indicate that only the 11.3~\um~PAH feature was detected, giving a range in possible star formation rates. BH masses are calculated both from the [OIV] linewidth and the $L_{bulge}-M_{BH}$ relation of \citet{ben09}. Errors on $M_{BH}^{bulge}$ are $\sim0.1$~dex.} 
\end{deluxetable*}

\subsection{Dust and gas temperatures}\label{ssec-temps}

We used mid-IR photometry from the IRS peakups and published IRAS fluxes \citep{san03} to fit a modified blackbody curve \citep{yun02} which fits for a dust temperature ($T_{dust}$) and an angular size (Table~\ref{tbl-derived}). Since the CSOs observed in 2009 were undetected by the IRAS satellite, fits for these objects are based only on the two IRS peakup points and are too undersampled for a measurement of $T_{dust}$. \citet{yun02} estimate the scatter on $T_{dust}$ in their sample to be $\sim10$K, while the formal errors on the curve-fitting give errors on the order of a few K. 

Out of the four CSOs with IRAS fluxes, 4C~+12.50, 4C~+31.04, and OQ~208 are moderately well-fit by a single-temperature blackbody, with $T_{dust}$ ranging from 65--110~K. The data for PKS~1413+135 has significant flux excesses to the fitted curve at higher energies; possible causes include contributions from hotter dust (500--2000~K), significant line emission in the 12 and 16~\um~bands, or non-thermal emission. This last option is supported by identification of PKS~1413+135 as a BL~Lac object, with beaming in the core contributing to continuum-dominated NIR and MIR spectra \citep{per96}. The blackbody fit for 1146+59 shows flux excess above the fitted curve at lower energies, which may indicate large amounts of dust at cooler temperatures $(<50$~K). As a sample, the relatively poor fits to a modified blackbody most likely indicate that these galaxies are complex systems without a single characteristic $T_{dust}$. \citet{deo09} show that Seyfert galaxies have at least three thermal components at $T\sim1000$, 200, and 60~K. 

\begin{figure*}
\includegraphics[scale=1.0]{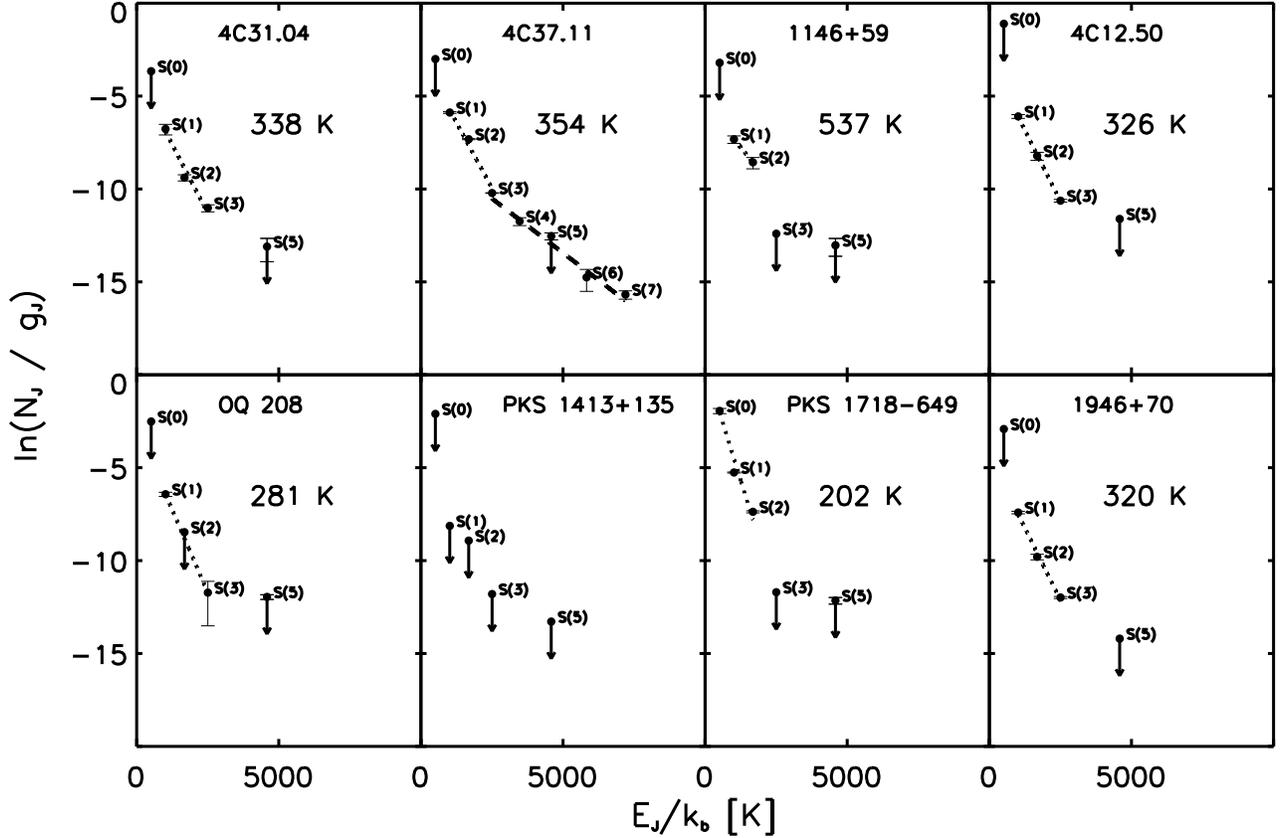}
\caption{\htwo~excitation diagrams for the CSO sample. Dotted lines show the $T_{ex}$ fit to the warm \htwo~gas, while dashed lines show the fit to a hot component (where detected). S(5) has error bars when the blended \htwo/[Ar~\scriptsize{II}\footnotesize] feature is detected, but has only an arrow when it is a true upper limit.}\label{fig-h2temp}
\end{figure*}

For the galaxies with multiple transitions of molecular hydrogen, we also fit excitation temperatures ($T_{ex}$) to a warm \htwo~component following the methods of \citet{rig02} and \citet{hig06}. Temperatures are calculated by assuming a Boltzmann distribution for the series of lines, the ratios of which depend only on $T_{ex}$; galaxies in which lines higher than S(3) are detected are typically better fit with a two-temperature model. The CSOs span a broad range of temperatures, with $T_{ex}$ ranging from 202~K in PKS~1718$-$649 to 537~K in 1146+59 (Figure~\ref{fig-h2temp}). The line luminosities can also be used to measure masses of the \htwo~components (Table~\ref{tbl-derived}); the mass of the warm gas varies over more than two orders of magnitude for the CSOs, with the smallest contribution coming from the gas in 1146+59 ($M_{H_2}=9\times10^4$~M$_\odot$) to the much larger component in the galaxy 4C~+12.50 ($M_{H_2}=4\times10^7$~M$_\odot$). 

4C~+37.11 is the only galaxy for which a hot gas temperature can be measured, at $T=854$~K; the derived hot gas mass is $\sim10\%$ of the warm \htwo~gas. Measurements of \htwo~from \citet{hig06} and \citet{rou07} combined with CO measurements suggest that the warm/hot components probed by the mid-IR lines are typically a small fraction of the total molecular gas mass ($1-30\%$), the majority of which exists at much lower temperatures. 

The excitation temperatures measured for the warm \htwo~gas are typically a factor of several higher than the dust temperatures measured by fitting the IR photometry. \citet{hig06} found no tendency for the \htwo~temperatures to depend on dust temperatures, based on IRAS fluxes. Results from the SINGS sample suggest that much of the \htwo~emission comes from either photodissociation regions (PDRs) associated with \HII~regions in star-forming galaxies, or from nuclear regions close to low-level AGN with an additional \htwo~component from shock heating \citep{rou07}. In both cases, the gas and dust are not expected to be co-spatial, and therefore temperature differences in a multi-phase medium are not unexpected.  


\subsection{Silicates}\label{ssec-silicates}


\begin{figure}
\includegraphics[scale=0.5]{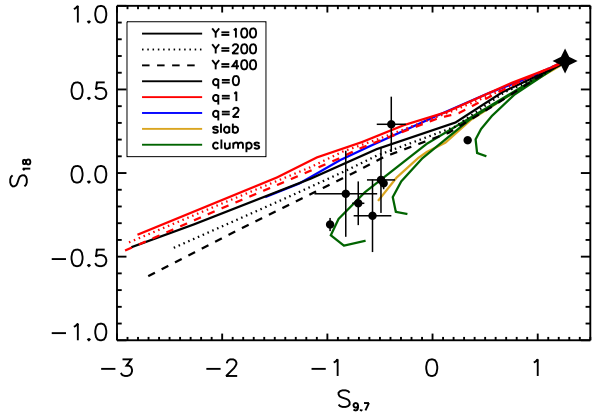}
\caption{Feature-feature diagram displaying the relative strengths of the 9.7 and 18~\um~silicate features. The lines represent modeling results of different dust geometries for cool, oxygen-rich silicates (OHMc) \citep{sir08}. Dotted, dashed, and solid lines represent the thickness of the dust shell in the spherical models ($Y=R_{outer}/R_{inner}$), and $q$ is the power law index for the density distribution of the shell ($\rho[r]\propto r^{-q}$). The green tracks show different numbers of dust clouds located along the line of sight for the clumpy model ($N_0$ = 1, 3, and 5 from upper right to lower left). The black star in the upper right corner is the starting point for all optically thin dust in the models ($S_{9.7}=1.26,S_{18}=0.67$). The CSOs are shown in black circles with error bars (smaller than the symbol in several cases).  \label{fig-feature2}}
\end{figure}

\citet{sir08} recently showed that the depths of the 9.7 and 18~\um~silicate features depend sensitively on the geometric distribution of the dust. They identify two distinct classes of ULIRGs, seen both in IRS data and in numerical models that distinguish between dust that is clumpy (AGN with broad optical lines) and smooth (geometrically and optically thick, with lower amounts of ionizing radiation). Plotting the relative strengths of the dust features against each other yields a "feature-feature" diagram (Figure~\ref{fig-feature2}), characterized by two regions representing the clumpy and smooth families in the optically thick limit. All Type 1 AGN in the sample of \citet{sir08} are found only in the clumpy region; LINERs and \HII~regions generally show deeper absorption at 9.7~\um, lying along the smooth track. Type 2 AGN may belong to either family. 

Seven of the CSOs lie along the track for clumpy dust. Six of these galaxies have silicate ratios with $N_0=3-5$ (the average number of dust clouds along a line of sight). The other galaxy lying in the clumpy region, OQ~208, is the only object in the sample showing the 9.7~\um~feature in emission, indicating that the dust is optically thin. Its location in Figure~\ref{fig-feature2} indicates that it has only $N_0\sim1-2$. In the standard AGN model, this could be explained by viewing a central power source and associated dust torus close to face-on, with a high inclination angle reducing the possibilities for absorption along the line of sight to the observer. This agrees with the optical classification of OQ~208 as a broad-line radio galaxy.

The silicate ratio for PKS~1718$-$649 is consistent with the \citet{sir08} models for smooth, geometrically thick spherical shells of dust. The galaxy has an optical depth of $\tau_V\sim100$; there are few ULIRGs from the \citet{spo07} and \citet{ima07} samples nearby in the feature-feature diagram. The host galaxy has an ``incomplete'' \HI~ring and trailing spiral arms, both of which argue against a smooth distribution. If so, this may indicate that the absorbing dust exists on smaller scales (less than 1~kpc) close to the center of the galaxy. 

We tested the robustness of the feature-feature diagram by running numerical models on the CSOs to quantify their physical parameters. We used the CLUMPY code to model a toroidal distribution of point-like dust clouds around a central source \citep{nen08,nen08a}. Individual clouds have optical depth $\tau_V$ and an exponential distribution such that the number of clouds per radial ray ($N_0$) is described by a Gaussian with angular width $\sigma$ in the polar direction. The thickness of the torus, $Y$, is scaled to the inner radius set by the dust sublimation temperature $(Y=R_{outer}/R_{inner})$. The final input parameters are the inclination of the disk $i$ and the power law index $q$ for the radial cloud distribution. The spectra are dominated by reprocessed dust emission independent of the heating source, but the AGN emission is visible for models with low optical depths and high inclination angles. 

For the CSOs, we sampled a coarsely-defined grid of 40,000 models across all six parameters. To assess the goodness-of-fit, we rebinned the IRS LR spectrum to the same wavelengths as the CLUMPY model and minimize the error $E$ from \citet{nik09}. This fitting also includes a multiplicative scaling of the spectrum accounting for the total bolometric flux of the AGN. 

\begin{figure}
\includegraphics[scale=0.5]{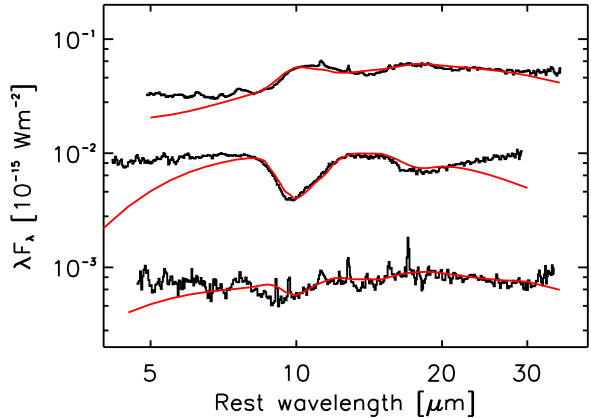}
\caption{IRS low-resolution spectra of three CSOs ({\it top to bottom:} OQ~208, PKS~1413+135, and 1946+70) with the best-fit models from the CLUMPY code overlaid in red. The minimized errors for the spectral fits are $E_{min}=1.52$, 0.50, and 0.20, respectively. Parameters of the CLUMPY fits are discussed in \S\ref{ssec-silicates}. \label{fig-clumpy}}
\end{figure}

Since CLUMPY only models the emission from blackbody dust, the fit is poor if the galaxy has strong emission features from atomic, molecular, or PAH features. In the CSO sample, the best fit models have $E_{min}$ between 0.20 and 1.52 - we show examples of three fits to the CSOs in Figure~\ref{fig-clumpy}. The best fit is for 1946+70, with $Y=100$, $N_0=10$, $q=1.0$, and $\tau_V=10$. The model reproduces absorption in both the 10 and 18~\um~features, but underpredicts emission shortward of 6~\um. This is a general trend for most CSOs in our sample, possibly due to hot dust ($T\sim1000$~K) or PAH emission. Neither $\sigma$ nor $i$ are well-constrained for any set of models. 

Two other CSOs show interesting fits from the CLUMPY models. OQ~208 exhibits the worst formal fit in the sample, with $E_{min}=1.52$. This is almost entirely due, however, to the underprediction of flux shortward of 9~\um~by $\sim0.1$~dex. The model fits the spectrum well between 9--28~\um, tracing both silicate features in emission. The best-fit models for this galaxy favor $N_0$ between 5 and 20, $q=2.0$, $\tau_V$ between 200 and 300, and $Y$ between 30 and 100. These parameters agree with general predictions of \citet{nen08a} and its placement in Figure~\ref{fig-feature2} with the exception of an unusually high $\tau_V$, rarely seen in Seyfert galaxies.  

PKS~1413+135 is another galaxy of particular interest due to its unusual spectrum - the lack of any atomic or molecular features make it an ideal candidate for radiative transfer modeling. The best-fit CLUMPY model for PKS~1413+135 has $Y=10$, $N_0=10-20$, $q=0.0-1.0$, and $\tau_V=10-30$. The model fits both dust absorption features well, but severely underpredicts the flux on either side of the silicate features (more than 0.5~dex at 5 and 30~\um). Both $q$ and $\tau_V$ agree with its placement in Figure~\ref{fig-feature2}, but the torus thickness $Y$ is unusually thin. The excess at the blue end could be due to additional reservoirs of hot dust not modeled by CLUMPY; however, the blue excess would likely require large amounts of very cold dust not supported by the amount of star formation seen. The best-fit model of PKS~1413+135 with a smooth spherical geometry yields $Y=100-500$, $q=1.5$, and $\tau_V=60-90$; this fits the blue excess well, but fails to account for the 18~\um~absorption and the red excess. The optical depth values are similar to values obtained from both the X-ray spectrum (Perlman et al., in prep) and the \HI~21-cm absorption line \citep{car92}. \citet{per02} propose a geometry with a single GMC lying along the line of sight, which agrees with the low $Y$ seen in our model. 

The individual CLUMPY models broadly support the conclusions from the feature-feature diagram. A clumpy dust geometry is a reasonable fit for most models, although there are clear contributions from dust at cold temperatures that are not properly modeled. PKS~1718$-$649, the only galaxy predicted to have a smooth dust shell, does not give a good fit to either silicate absorption feature with the CLUMPY geometry. 


\subsection{Quantifying the AGN fraction}\label{ssec-agnfraction}


\begin{figure}
\includegraphics[scale=0.5]{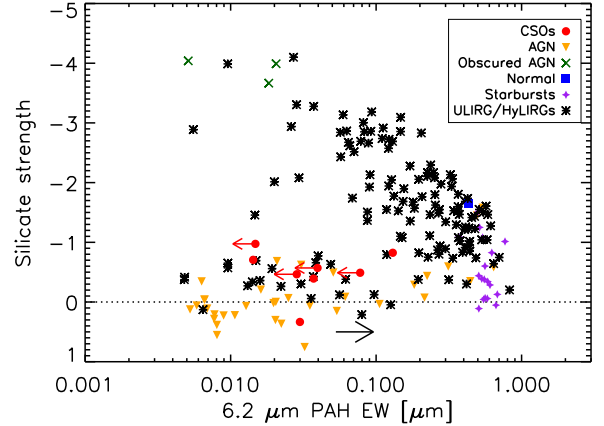}
\caption{``Fork'' diagram displaying the 6.2~\um~PAH equivalent width vs. the silicate strength at 9.7~\um. CSO galaxies are shown in red; additional \textit{Spitzer} data are from \citet{spo07} and Willett et~al. (2010). Limits are given on the PAH EW for CSOs with no measured flux above the noise. The black arrow gives the mean effect on the PAH EW if fluxes from PAHFIT are used in place of the spline-fit technique. \label{fig-fork}}
\end{figure}

A powerful mid-IR diagnostic for separating populations of starburst galaxies from AGNs is the ``fork diagram'' developed by \citet{spo07}, which plots the 6.2~\um~PAH equivalent width (EW) against the depth of the 9.7~\um~silicate absorption. Two distinct families are seen on this plot, with a horizontal branch of low silicate absorption/emission and variable PAH content largely populated by AGN, and a diagonal branch with deeper dust absorption and stronger PAH emission populated mostly by starburst galaxies. We overlay the data from the CSOs on the fork diagram containing a larger sample of galaxies in Figure~\ref{fig-fork}. 

4C~+31.04 lies between the two branches, with a PAH EW that does not clearly classify it as either a starburst galaxy or an AGN. The remaining CSOs in the sample lie along the horizontal branch in the locus of AGN galaxies, while four have upper limits on the PAH EW that may place them even further from the starburst branch. 

The separation of galaxies into two distinct branches is related to the geometric distribution of the dust, as shown in Figure~\ref{fig-feature2}. The horizontal branch in Figure~\ref{fig-fork} is defined by the low amounts of 9.7~\um~absorption found in clumpy models, with the varying PAH EW a function of either the ionizing flux from an AGN (which dissociates PAHs) or the metallicity of the galaxy. The diagonal branch contains the majority of the galaxies with deep silicate absorption and requires a geometrically smooth distribution; while this can harbor an AGN at the highest levels of obscuration, none of the CSOs are located in this region. This suggests that none of these CSOs have large amounts of star formation often seen in ULIRGs. 

\begin{figure}
\includegraphics[scale=0.5]{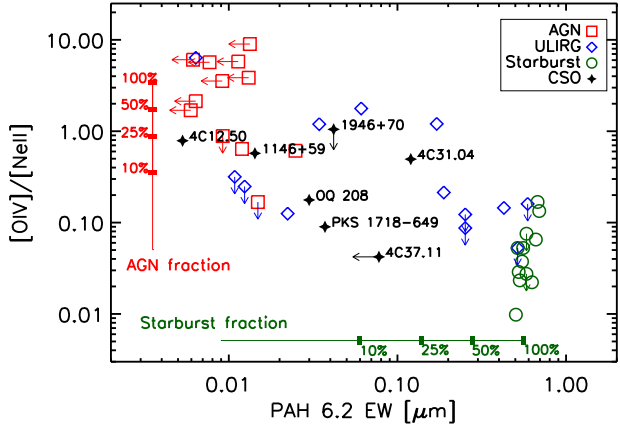}
\caption{The PAH 6.2~\um~equivalent width vs. the flux ratio of the [O~\scriptsize{IV}\footnotesize]~25.9~\um~and [Ne~\scriptsize{II}\footnotesize]~12.8~\um~transitions. We compare the CSOs to samples of AGN, ULIRGs, and starburst galaxies from \citet{arm07}. The mean of the AGN ({\it red}) and starburst galaxies ({\it green}) are assumed to set the 100\% level of AGN and star formation contributions, respectively - the bars show the fractional contribution assuming simple linear mixing. \label{fig-pah-oiv}}
\end{figure}

A second method of assessing the AGN contribution uses the 6.2~\um~PAH EW plotted against the \oIV/\neII~ratio (Figure~\ref{fig-pah-oiv}). Since \oIV~is known to arise from both AGN and star formation, comparing the two addresses the source of these moderate-ionization lines. For comparison, we use data from \citet{arm07} where we assume that the \oIV~emission from a pure AGN represents 100\% of the contribution to $L_{IR}$; similarly, we use a sample of pure starburst galaxies to approximate the contribution from star formation. A simple linear mixing is assumed for tracing the distribution between the extremes. 

Two CSOs (4C~+12.50 and 1146+59) fall in the AGN family, with $<2\%$ of the expected PAH contribution from star formation. PKS~1413+135 has no \oIV~or \neII~lines, but a PAH EW upper limit of $<0.1$~\um. The remainder show a larger star formation contribution to the \oIV~emission, with the highest starburst fraction at $\sim25\%$. Since no CSO shows either AGN or star formation contributions above 30\% from this diagnostic, this indicates that CSOs form an intermediate class with substantial contributions from both energy sources. 

\begin{figure}
\includegraphics[scale=0.5]{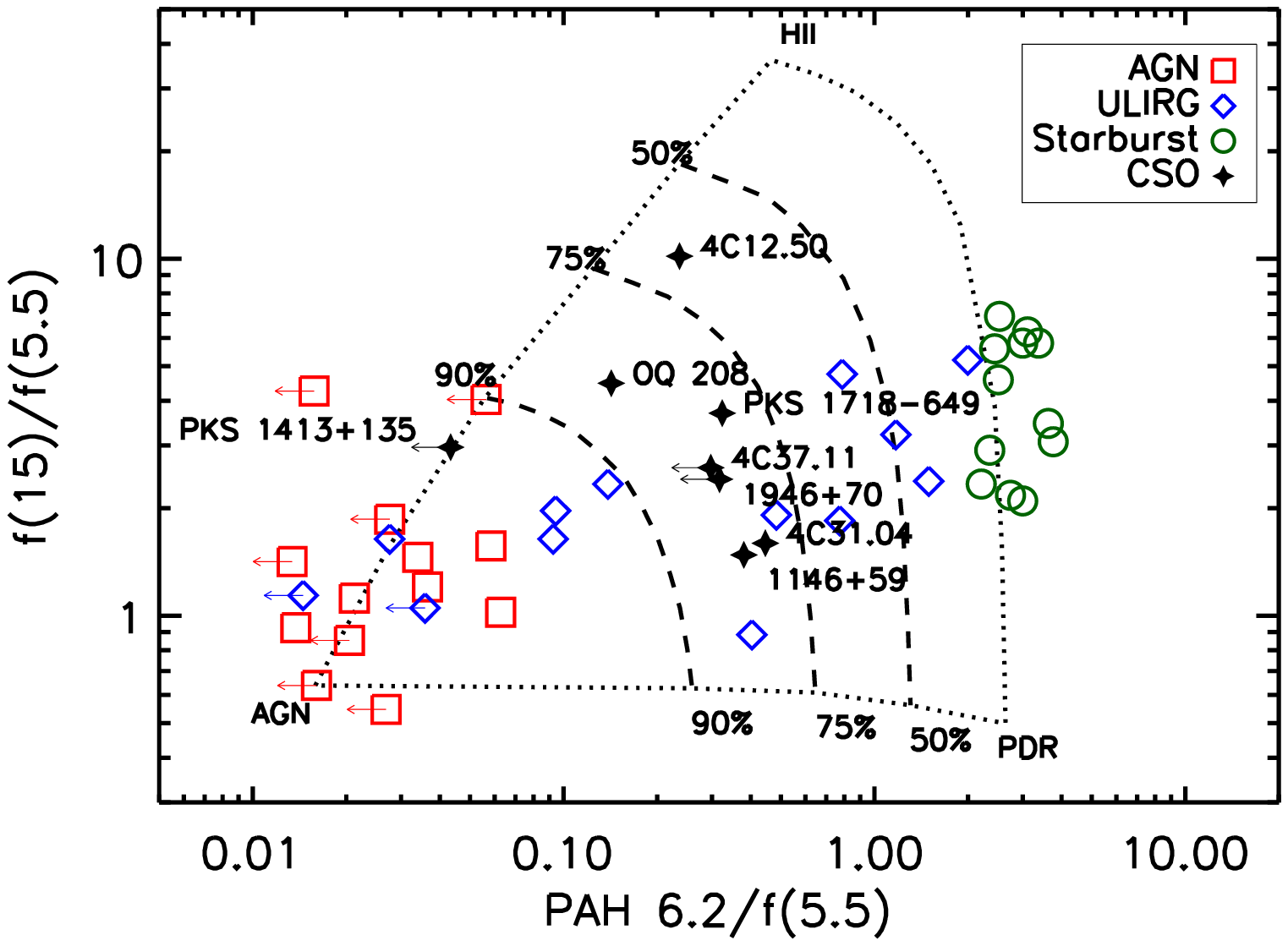}
\caption{Peak PAH flux at 6.2~\um~(normalized to the 5.5~\um~continuum) vs. the flux ratios at 15 and 5.5~\um~\citep{lau00}. The PAH emission traces star formation, while the f(15)/f(5.5) ratio is inversely proportional to the amount of hot dust in the galaxy. Symbols are the same as in Fig.~\ref{fig-pah-oiv}. The dotted line vertices represent ``pure'' examples of an AGN (3C~273), an H\scriptsize{II}\footnotesize~ region (M17), and a photo-dissociation region (PDR; NGC~7023); dashed lines indicate the fractional AGN contribution. \label{fig-triangle}}
\end{figure}

Finally, we compare the star formation contribution to the amount of hot dust present in the galaxies. AGN can heat silicate dust to temperatures in excess of $1000$~K, increasing the continuum emission at shorter wavelengths \citep[e.g.,][]{pie92,alo01}. Figure~\ref{fig-triangle} compares the PAH strength to the flux ratio at 15 and 5.5~\um, which will be lower as the amount of hot dust increases. The three vertices represent "pure" examples of an AGN, an \HII~region, and a photo-dissociated region (PDR), along with data for AGN, starburst galaxies, ULIRGs, and CSOs. PKS~1413+135 is the only CSO that falls strongly in the AGN region - the remainder of the galaxies are scattered with ULIRGs indicating contribution from both star formation and AGN heating. The scatter between the two families is consistent with the trends seen from both the silicate depths and \oIV~ratios (Figs. \ref{fig-fork} and \ref{fig-pah-oiv}). There is no obvious correlation of the AGN fraction with silicate depth, kinematic ages of the jets, or optical type. 

The mix of AGN and star formation activity in all three diagnostic plots supports the picture that the active radio phase may be triggered following a merger. The $10^8$~yr delay seen in CSOs by \citet{per01} provides enough time for a burst of star formation triggered by the initial pass; afterwards, as gas is driven to the center of the merging galaxies, the AGN phase begins to dominate the IR emission. The CSO outburst, however, appears to take place well before the AGN overwhelms star formation as a photon source in the IR. The scatter in AGN fraction in the CSOs may have several sources, including differences in viewing geometry, the total amount of star-forming gas, and the time since the beginning of the active phase. 

\subsection{Star formation rates}\label{ssec-sfr}


\begin{figure}
\includegraphics[scale=0.5]{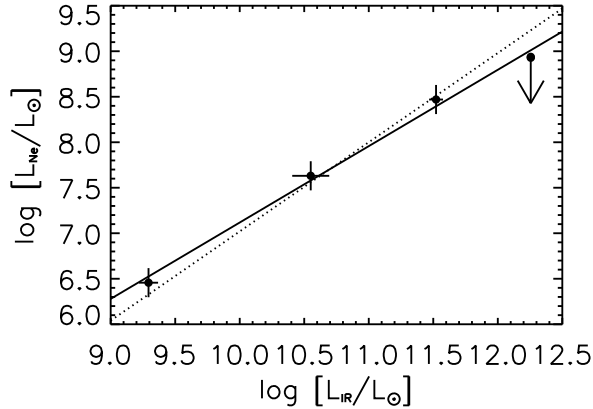}
\caption{Luminosity of [Ne~\scriptsize{II}\footnotesize] + [Ne~\scriptsize{III}\footnotesize] lines as a function of $L_{IR}$ for the four CSOs with IRAS detections. Solid lines show the best linear fit for the CSOs; the dotted line shows the fit for the much larger sample of \citet{ho07}. An upper limit is given for PKS~1413+135, which shows no detection of either neon line.\label{fig-sfr_cso}}
\end{figure}

We quantify star formation rates (SFR) using the relation from \citet{ho07}, where the intensities of mid-IR neon emission show a tight linear correlation with both $L_{IR}$ and the ionizing Lyman-$\alpha$ continuum from young stars. The combined luminosity from both the \neII~and \neIII~lines are used to determine the SFR:

\begin{equation}
\label{eqn-sfrne}
{\rm SFR}~[M_\sun/{\rm yr}] = 4.73\times10^{-41}\left[L_{NeII}+L_{NeIII}\right],
\end{equation}

\noindent where the neon luminosities are measured in erg~s$^{-1}$. Following \citet{ho07}, we assume $f_{ion}=0.6$ (the fraction of ionizing photons trapped by the gas) and that the fractions of singly- and doubly-ionized neon are $f_+=0.75$ and $f_{++}=0.1$. We compare this calibration for ULIRG data to the $L_{IR}$-$L_{Ne}$ fit for the CSOs (Figure~\ref{fig-sfr_cso}). IRAS fluxes are available for four CSOs in the sample, which we use to compute $L_{IR}$ \citep{san96}; the fit is reasonably good for all objects, even with the CSOs varying in $L_{IR}$ by more than three orders of magnitude. 


\begin{figure}
\includegraphics[scale=0.5]{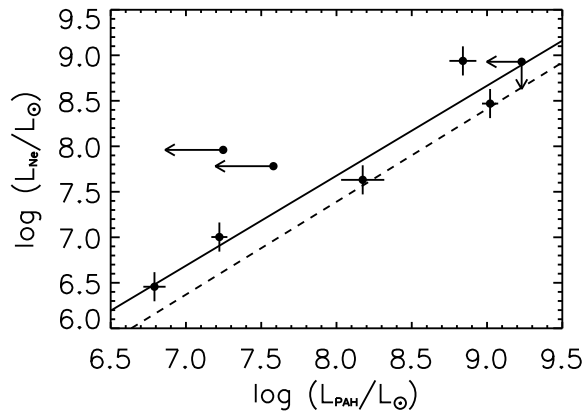}
\caption{Combined luminosity of the 6.2 and 11.3~\um~PAH features as a function of the [Ne~\scriptsize{II}\footnotesize] + [Ne~\scriptsize{III}\footnotesize] luminosities. The dashed line shows the linear fit for the PAH-neon relation calibrated for ULIRGs \citep{far07}; the solid line shows the linear fit to the CSOs with neon detections.\label{fig-cso_nepah}}
\end{figure}

\citet{far07} use a second method of determining star formation rates that makes use of the observed linear correlation in ULIRGs between the luminosities of \neII+\neIII~and the 6.2 and 11.3~\um~PAH features:

\begin{equation}
\label{eqn-sfrpah}
{\rm SFR}~[M_\sun/{\rm yr}] = 1.18\times10^{-41}~L_{PAH}~[{\rm erg~s}^{-1}].
\end{equation}

\noindent A positive correlation between the PAH and neon emission is also observed for CSOs (Figure~\ref{fig-cso_nepah}). While a slope fitted to the data is consistent with a linear relation ($1.0\pm0.4$), the scatter is nearly an order of magnitude larger than that measured for the ULIRG sample ($1.02\pm0.05$). This may be an indication that some CSOs have neon emission that does not arise from recent star formation. Deviations from the linear relation also are expected for low-metallicity galaxies \citep{pee04,cal07} for which both neon and/or PAH features can be significantly depleted with respect to the true SFR. 4C~+12.50, in particular, deviates significantly both from the ULIRG fit and the remainder of the CSO sample (showing almost an order of magnitude less PAH emission than would be expected from the neon strengths).  

Star formation rates in CSOs vary by more than two orders of magnitude over the sample (Table~\ref{tbl-derived}), with the lowest measured rate in 1146+59 ($0.3-0.5M_\sun$/yr) and the highest in OQ~208 and 4C~+12.50 ($<45M_\sun$/yr). The SFR measured from the PAH strengths is consistently lower than the SFR from neon by a factor of 4--5; however, the conversion between neon and PAH emission depends on a number of assumptions (including the timescale of star formation and the conversion to Lyman continuum flux). In addition, the data used to calibrate Equation~\ref{eqn-sfrpah} is based on data from ULIRGs, with unclassified contributions to their active phases; this may contribute to the scatter in Figure~\ref{fig-cso_nepah}. While emission from AGN-heated gas can contribute to neon emission (which may explain the large disparity of the SFRs for 4C~+12.50), PAHs are easily dissociated by hard radiation environments and are almost never detected in AGN. For this reason, we regard PAHs as the more robust SFR indicator; this also implies that the majority of the star formation being detected takes place in the outer regions of the CSOs. 

\subsection{Black hole masses}\label{ssec-bhmass}

High-ionization transitions in the mid-IR that are assumed to emit from narrow-line regions gravitationally bound to an AGN can be used to estimate the mass of the central black hole \citep{das08}. For the six CSOs in which the \oIV~25.8~\um~transition is detected, the line is spectrally resolved for two objects: 1146+59 and PKS~1718$-$649. The \oIV-$M_{BH}$~calibration yields black hole masses of log($M_{BH}/M_\sun)=9.2$ and $8.6$. 4C~+31.04 has a measured \oIV~dispersion above the instrumental resolution but not satisfying the resolution criterion of \citet{das08}; if valid, it has a SMBH mass of log($M_{BH}/M_\sun)=8.2$. The velocity dispersions of two other CSOs in which \oIV~was detected, OQ~208 and 4C~+37.11, are below the resolution of the HR module and give only upper limits on a central black hole mass (Table~\ref{tbl-derived}). 

The \oIV~line is also detected in 4C~+12.50 with a FWHM of $1026\pm97$~km/s, which would correspond to a central BH mass of log($M_{BH}/M_\sun)=11.5$ if caused entirely by the black hole. This is more than two orders of magnitude larger than the maximum BH mass used to calibrate the relation of \citet{das08} and is significantly higher than predicted upper limits on ultramassive black holes. 4C~+12.50 has two bright optical nuclei within a common envelope separated by only 1.8$\arcsec$ \citep{gil86,axo00}, which is several times smaller than the width of the LH slit. The broad linewidth is likely due to emission from both nuclei moving at relative velocities of a few hundred km/s. Several other lines in this object, including \sIV~and \neIII, also have FWHM in excess of 1000~km/s. The \oIV~line in this CSO thus cannot be used to estimate an accurate central BH mass. 

Since \oIV~can also be produced in extreme starbursts, the measured line dispersions could represent motions of the gas in the ionized environments surrounding young, hot stars or the relative motions between star forming regions. The star formation rates from 1146+59 and PKS~1718$-$649, however, are the lowest measured in the sample ($<2$~M$_\sun$/yr, from both neon and PAH data). If star formation is not the main contributor to \oIV~emission, then the lines are more likely caused by emission from a narrow-line region. Uncertainties in the physical location of the ionized gas with respect to the black hole, however, contributes to the uncertainty.  

We use a second method to estimate the black hole masses from the well-known bulge-luminosity relation. We assume that the total luminosity of the galaxy comes from the bulge (which could be an overestimate for spiral hosts or mergers with extended features). We use $g$ and $r$ photometry from the SDSS for five CSOs to calculate $L_V$; for the remaining galaxies not in the SDSS, we use H-band NICMOS and V-band POSS measurements and apply typical colors for an elliptical galaxy. We use the AGN-calibrated $M_{BH}-L_{bulge}$ relationship from \citet{ben09} to compute central BH masses; masses for all eight CSOs lie between log~$M_{BH}=8.0-9.0$. The measurement for PKS~1718$-$649 agrees well with the mass from the \oIV~dispersion, but the bulge $M_{BH}$ for 1146+59 is lower than the \oIV~estimate by $\sim1$~dex. 

\subsection{Ionization and excitation states}\label{ssec-excitation}


\begin{figure}
\includegraphics[scale=0.5]{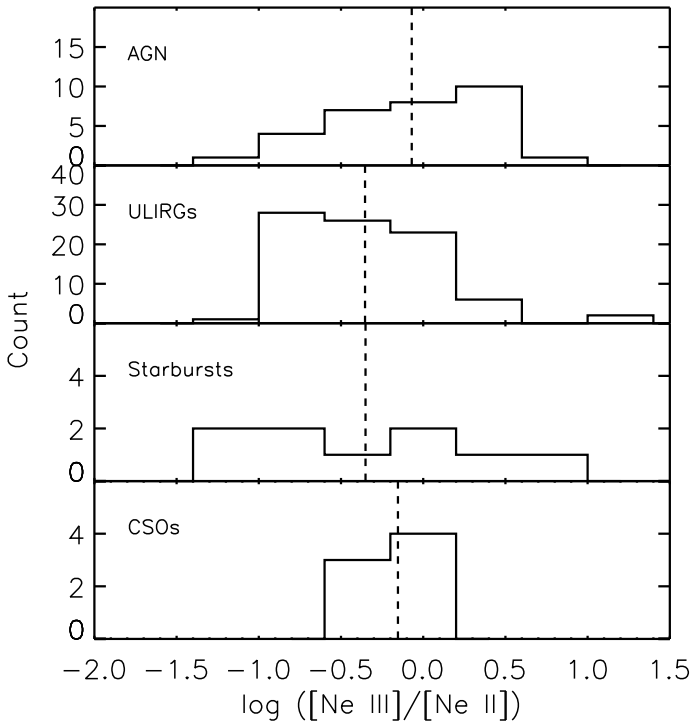}
\caption{Distribution of the [Ne~\scriptsize{III}\footnotesize]/[Ne~\scriptsize{II}\footnotesize] line ratio; higher ratios trace a harder ionizing environment. For comparison, we also show neon ratios for samples of known ULIRGs (Farrah et al. 2007; Willett et al. 2010), AGN \citep{stu02,tom08}, and starburst galaxies \citep{ver03}. Dashed lines show the mean value for each class of objects. }\label{fig-excitation}
\end{figure}

The hardness of the ionizing radiation field in the 300--600~\AA~range can also be traced using the excitation states of fine-structure lines from the same species; a harder radiation field increases the line ratio of higher/lower states. In the mid-IR, the \neIII/\neII~ratio is the only pair of lines visible in a majority of the CSOs. We plot the line ratios for the eight objects in which neon is detected (Figure~\ref{fig-excitation}), along with the distribution of other IR-bright galaxies measured with the ISO and {\it Spitzer}. 

The neon ratios in all CSOs are between 0.4 and 1.0. The average neon ratio from the AGN samples of \citet{stu02} and \citet{tom08} is $1.3\pm1.0$. Starbursts with mid-IR data \citep{ver03} are much fewer in number and show a larger spread in neon ratios, with a mean value of $1.8\pm3.3$. 

The distribution of CSO neon ratios is not significantly different from either starbursts or AGN, although comparisons are limited due to the small number of CSOs. The mean hardness ratio in AGN is higher than that in the wider ULIRG population, likely due to the harder X-ray/UV environment that can arise from accreting gas around a central black hole. Given the large overlap in hardness ratios for both AGN and starburst galaxies, this parameter alone cannot classify CSOs in either category. 

\subsection{Mid-IR classification of the CSO hosts}\label{ssec-classification}

The mid-IR spectra show a heterogeneous set of features for the CSOs, as discussed above. While the objects were selected based on similar characteristics in the radio, analysis of the mid-IR features shows significant deviations from a single CSO ``family''. We classify the objects into four categories, based on features visible in the LR spectra (PAH strength, silicate depths, fine-structure and \htwo~lines):

\begin{list}{$\bullet$}{}
 \item Category 1 shows several weak-to-moderate PAH emission features, often with only an upper limit on the 6.2~\um~feature, with moderate $S_{9.7}$ absorption. This is consistent with low levels of star formation and a moderately dusty environment. This is the most common category for the CSOs observed, including 4C~+31.04, 1146+59, 4C~+12.50, PKS~1718$-$649, and 1946+70. 
 \item Category 2 is similar to objects in Category 1, with weak PAH emission and moderate $S_{9.7}$ absorption; however, these objects display strong \htwo~features that may arise from shocked molecular gas. 4C~+37.11 is the lone example of this in the sample. 
 \item Category 3 also has weak PAH emission on top of a strong continuum, but shows the silicate dust features in emission rather than absorption. This is a strong signature of a quasar or AGN in the galaxy \citep{hao07}, although active star formation may still be taking place. OQ~208 is the only object in this category. 
 \item Category 4 objects have {\em no} discernible emission features, with the only break in a continuum that can be fit with a single power law the silicate absorption at 9.7 and 18~\um. PKS~1413+135 is the only CSO in the sample with these basic blazar characteristics. 
 \end{list}

The IRS non-detection of the CSO 1245+676 (\S~\ref{ssec-no1245}) is also interesting. The object is detected with 2MASS with $m_{K_s}=13.2$, suggesting the existence of a large population of old, red stars. The low levels of mid-IR emission, however, imply that the object is dust-poor and has no strong PAH features which would signal high levels of star formation. These observations are consistent with the standard picture of an AGN host as an elliptical galaxy with little gas or dust. The host galaxy of 1245+676 thus might be an example of a dust-poor galaxy similar to Categories 1 or 2; deeper observations that can place strong limits on the PAH emission and silicate absorption are critical to answering this question. 

In summary, 5/8 CSOs have moderately dusty environments and moderate levels of star formation. The exceptions from this typically show strong evidence for being AGN-dominated, including the broad-line galaxy OQ~208 and the BL~Lac PKS~1413+135. 


\section{Discussion}\label{sec-discussion}

\subsection{Photoionization vs. star formation}\label{ssec-photoionization}

One of the key issues in interpreting the mid-IR data is the ionization source of the emission features. For IR-bright galaxies, this typically has two main possibilites: photoionization of the gas by an AGN, or from star formation possibly spread throughout the galaxy. If the former is the case, then it is likely that some of the gas is close enough to accrete onto the AGN and provide a power source for the radio jets in the inner kpc.

For several CSOs, hard X-ray emission provides direct evidence for supermassive BHs; in the case of 4C~+37.11, a pair of BHs has been directly detected. In the optical, complex Balmer profiles in OQ~208 also identify a broad-line region surrounding an AGN. The fork diagram places all eight CSOs on the AGN branch, with the silicate absorption ratios favoring a clumpy dust torus for most galaxies in the sample. Finally, multiple CSOs show very low levels of star formation (from multiple tracers). PKS~1413+135 has only upper limits on star formation from mid-IR tracers, although the limits are higher than most SFR measurements in the sample due to its much greater distance. 

There is also evidence that star formation may be responsible for a significant fraction of the mid-IR emission, however. Two CSOs (4C~+12.50 and OQ~208) show strong PAH features that would normally be dissociated if directly exposed to an AGN, with $L_{IR}$ and $L_{Ne}$ also supporting $SFR>50M_\sun$/yr in both galaxies. The detection of \htwo~in most galaxies indicates the presence of molecular gas - this has also been detected at cooler temperatures at mm-wavelengths in several CSOs, meaning that star-forming gas is still present in most of the sample. \neV, a ``smoking gun'' of AGN activity in the mid-IR, is not detected in any CSO, while the \oIV~emission could arise from photoionization from either O and B stars or an AGN. Finally, the poor fits of many CSOs to the classic AGN torus of clumpy dust could indicate a different geometry, with distributed dust and gas undergoing star formation in extended features. 

The results of the mid-IR data also have intriguing implications for the evolutionary history of radio galaxies. Although the sample size is small, nearly all CSOs show contributions from both AGN and star formation (Figs.~\ref{fig-fork}-\ref{fig-triangle}). This is strong support for the scenario in which young radio galaxies are formed by the merger of two disks; remnant star formation from the initial merger is still detected, but clear signatures of AGN have also begun to appear in other galaxies (\oIV, hot dust, and a clumpy dust geometry in the mid-IR; X-ray emission and broad-line regions also support this). Since we are deliberately selecting only the youngest radio galaxies, it is not surprising that the AGN do not yet dominate the IR emission if they are also in their earliest stages. As the galaxies continue to virialize (and possibly feed the central BH), we predict that star formation tracers such as PAH emission should become weaker, the AGN and hard radiation environment will increase, and the jets continue to advance, eventually forming FR~II-type radio galaxies \citep{beg96}. Mid-IR observations of later-stage radio galaxies (``medium symmetric objects'', or MSOs) could also support this hypothesis if they show a larger AGN fraction than their younger counterparts. 

\subsection{What is the power source for the radio jets?}\label{ssec-powersource}

Mid-IR data can discriminate between two quite different models for powering radio-loud AGN: accretion from a disk and/or torus as opposed to the tapping of spin energy from a supermassive black hole \citep{bla77,wil95}. If accretion is the energy source, then there should be evidence for enough dust and gas in the nuclear region to obscure the AGN in near-IR NICMOS images without obvious broad lines. Furthermore, we would expect to detect highly-ionized fine-structure transitions such as \neV \citep{arm04}. The slow Bondi accretion vs. ADAF models, however, are not so easily distinguished from mid-IR data.

Nearly $75\%$ of all QSOs show \neV~emission and $>90\%$ show \oIV~25.9~\um~emission indicative of accreting gas close to the black hole \citep{sch06}. The sample of \citet{arm04} also detected \neV~in 3/10 ULIRGs. If the absence of \neV~is a physical result (and not a question of sensitivity), then this requires a substantial energy source not powered by accretion. The non-detection of \neV~in all eight CSOs, based on simple Gaussian statistics for QSOs, is significant at the $\sim4\sigma$ level. 

The lower ionization potential transition \oIV~was detected in six CSOs; this line can be associated both with AGN and photoionization from extreme starbursts. Possible scenarios in which an AGN could produce \oIV~without \neV~are: significant depletion of neon in the accreting gas; a differential extinction screen that obscures the neon emission; or, a radiation environment in which the majority of the ionizing flux has energies between $55<E_{ion}<97$~eV. This may be the case in lower-luminosity AGN, compared to luminous QSOs. The limits on \oIV~and \neV~24~\um~for the most continuum-dominated CSOs (PKS~1413+135 and 4C~+12.50) are nearly an order of magnitude higher than the brightest detections. We also test the \neV~upper limits by comparing it to the average \neV/\neII~ratios in PG quasars, which are robust for 6/7 CSOs with \neII~detections and indicating that signal-to-noise ratio is not the primary factor in the absence of \neV. 

We can also address the possibility of accretion fueling the radio jets from an energy balance argument. The power in the radio jets has two components: the mechanical power necessary to excavate the radio cavity in the ISM and the radio luminosity from the jets themselves. Following \citet{all06}, the work needed to expand the cavity is:

\begin{equation}
\label{eqn-pjet}
P_{jet} = \frac{\gamma}{\gamma - 1} p V / t_{age}, 
\end{equation}

\noindent where $p=kT\rho/\mu m_p$ is the pressure of the surrounding ISM and $\gamma$ is the adiabatic index of the gas (assumed to be $5/3$). We assume a temperature of 1~keV and $n\simeq1$~cm$^{-3}$, based on the largely homogeneous results for elliptical galaxies in the sample of \citet{all06}. While the cavities cannot be directly observed without high spatial resolution X-ray images, we use the projected dimensions of the radio jets as an estimate. We assume $V = \pi r_l r_w^2$ is the volume of a cylindrical cavity with projected radii of $r_w$ and $r_l$ along the jet minor and major axes. We combine $r_l$ with the kinematic ages ($t_{age}$) of the CSOs from VLBI proper motions to determine the timescale necessary to excavate the cavity. 

We add $P_{jet}$ to the total radio luminosity assuming a self-absorbed synchrotron spectrum with the low-frequency emission $S\propto\nu^{5/2}$. The high-frequency spectral index is measured (where possible) using the 4--8~GHz fluxes. In almost all CSOs, we find that $P_{radio}$ is 2--3 orders of magnitude larger than $P_{jet}$, the opposite of jet-dominated FR~I galaxies. 

To test the accretion model, we calculate the Bondi power for spherically symmetric absorption:

\begin{equation}
\label{eqn-pbondi}
P_{Bondi} = 4 \pi \lambda G^2 M_{BH}^2 \rho \eta c^2 c_s^{-3}, 
\end{equation}

\noindent where $\lambda=0.25$ is a numerical coefficient depending on the adiabatic index of the accreting gas, $\eta$ is the efficiency of accretion, $c_s$ and $\rho$ are the sound speed and density of the gas, and $M_{BH}$ the mass of the central black hole onto which the gas is accreting \citep{bon52}.  The number density is consistent with the findings of \citet{pih03} requiring $n_{ISM}\sim1-10$~cm$^{-3}$ to confine GPS/CSS sources. The black hole mass is estimated using the bulge luminosity-$M_{BH}$ relation. 

\begin{figure}
\includegraphics[scale=0.5]{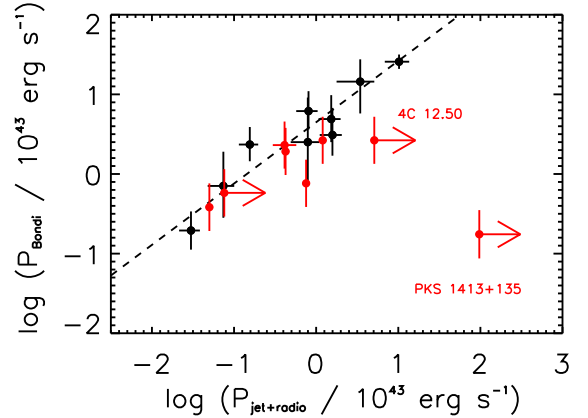}
\caption{Bondi accretion power as a function of the combined jet mechanical power and radio luminosity. The red dots are CSOs, while the black dots are X-ray luminous elliptical galaxies from \citet{all06}. Arrows show the limits on CSOs without constrained $t_{age}$. The dashed line is a linear fit to the X-ray ellipticals; we label CSOs that deviate significantly from this fit. }\label{fig-bondi}
\end{figure}

Models of jet power both in elliptical galaxies \citep{all06} and galaxy clusters \citep{bir04} demonstrate a scaling relation between Bondi accretion and the jet power. Results for CSOs are shown in Figure~\ref{fig-bondi}. For an assumed accretion efficiency of $\eta=0.1$, Bondi accretion is capable of powering the radio jets in most objects, with the prominent exceptions of 4C~+12.50 and PKS~1413+135. If advection-dominated accretion flows can be shown to exist without production of the high-ionization \neV~lines, then we must still consider gas accretion as a viable power source. 

As an alternative, we examine the evidence for the \citet{wil95} model in which the spin energy of two SMBHs is tapped to power the radio jets in the absence of accreting gas. Since the delay between the merger and the onset of radio activity and the timescale of merging black holes is both $\sim10^8$~yr. However, the general model offers relatively few testable predictions that can be made in the absence of accreting gas. One constraint, however, is that the BHs must be of comparable size and have masses greater than $10^8 M_\sun$. In the case of 4C~+37.11, these conditions have already been met; the BH masses measured both with $L_{bulge}$ and the \oIV~dispersion also lie above this line. The separation of the two BHs (7.3~pc), however, mean that the cores are still too distant to tap the spin energy for powering the radio luminosity. No radio-loud galaxy has yet been discovered in which merging BHs are close enough to tap their combined spin energy. 


\section{Conclusions}\label{sec-conclusion}

We observed a sample of eight low-$z$ CSOs with the IRS, representing very young radio galaxies and possible progenitors to FR~IIs. {\it Spitzer} investigations have revealed:

\begin{enumerate}
 \item The mid-IR SEDs for CSOs show significant variety. This includes silicate features seen in both emission and absorption, spectral indices indicating dust at a wide range of temperatures, and fine-structure and PAH emission. We sort the galaxies into four categories based on their IRS spectra, with the most common features including weak 9.7~\um~dust absorption, moderate PAH and atomic emission, and \htwo~temperatures of 200--400~K. There is no evidence in the mid-IR, however, of a single ``family'' for CSOs. 
 \item PKS~1413+135 is a unique object among CSOs, showing no emission or absorption features associated with gas in the ISM. The only identifiable features in the mid-IR are the silicate dust troughs at 9.7 and 18~\um. While the dust redshifts are roughly consistent with the near-IR obscuration, it does not address the possibility that the AGN lies along the line of sight and may not be physically associated with the optical counterpart \citep{per02}. 
 \item Numerical models indicate that a clumpy dust torus can fit the silicate features in most CSO spectra. Features common to most models have 1--10 dust clouds along the line of sight, with optical depths reaching as high as a few hundred in $\tau_V$. 
 \item Multi-wavelength evidence (including X-ray fluxes, silicate/PAH ratios, and radio-imaging of SMBHs) indicate that significant fractions of the sample have AGN at their centers. Evidence for continuing star formation, however, suggests that CSOs occupy a continuum between starburst galaxies and AGN. This supports the scenario that the elliptical hosts were formed by a merger of disk galaxies with a delay before the activation of an AGN.  
 \item Based on the work necessary to expand the radio jet cavities, accretion is a viable power source for 6/8 CSOs in the sample. The non-detection of high-ionization lines and absence of large extinction support an ADAF model, if this is the case. Although BHs with $M>10^8M_\sun$ likely exist in all CSOs, there is no direct evidence that BH spin energy provides the power source for the radio jets. 
\end{enumerate}


\acknowledgments

This work is based on observations made with the \textit{Spitzer Space Telescope}, which is operated by the Jet Propulsion Laboratory (JPL) and the California Institute of Technology. We also made use of the NASA/IPAC Extragalactic Database (NED) which is operated by JPL and Caltech under contract with NASA. Many thanks are due to L.~Armus, V.~Charmandaris, and H.~Spoon for their help with data reduction, K.~Dasyra and B.~Keeney for useful discussions, and to the Spitzer Science Center for hosting KW and JD while collaborating on data analysis. KW and JS gratefully acknowledge support from {\it Spitzer} programs 30515 and 50591. 


\begin{appendix}\label{appendix}

In this section, we summarize the multi-wavelength properties of the CSOs from the literature. Objects are referred to primarily by their radio nomenclature, rather than the host galaxy. \\

\noindent {\bf 4C~+31.04}: The radio emission has a compact core with symmetric hotspots and an unusual ``hole'' in its eastern lobe. \citet{gir03} measure lobe advance speeds of $(0.33\pm0.06)h_{65}^{-1}c$ and a kinematic age of $550\pm100$~years. The distance from the core to the hotspot edges is 50--60~pc for both lobes. The CSO's host galaxy is MCG 5-4-18, which has a nearly edge-on circumnuclear disk and strong obscuration in both the nucleus and outer regions. ``Cone-like'' features in the nucleus, seen both in \HI~\citep{con99} and HST images \citep{per01}, are aligned with the radio jets. \citet{gar07} detect a molecular/dusty disk around the AGN with $M_{gas}>5\times10^9~M_\sun$ and a size of 1.4~kpc. Distortions and non-circular motions in the molecular disk suggest it is not dynamically relaxed, possibly resulting from either a recent merger, gas accretion events, or interactions between the radio jet and the disk.  

\noindent {\bf 4C~+37.11}: This galaxy possesses several features highly unusual among CSOs. It has extended kiloparsec radio emission that may be either due to a merger or previous activity in one of the cores. The parsec-scale radio structure shows two compact, flat-spectrum sources identified as supermassive black holes (SMBHs) that have not yet coalesced \citep{man04}, with a combined mass of $1.5\times10^8~M_\sun$ and a separation of only 7.3~pc \citep{rod06}. The kinematic signature of the BH also appears in a pair of broad \HI~absorption features \citep{mor09}. The radio jets are 15--25~pc in length, with the north hotspot moving outwards at $(0.137\pm0.034)h_{71}^{-1}c$; this yields a kinematic age of $502\pm129$~years. The host galaxy is the elliptical 0402+379, which \citet{sti93} describe as having a ``flat brightness distribution'' with no central point source in the optical. The X-ray brightness of this galaxy is the highest in the sample, and it is the only known CSO detected in the {\it ROSAT} All-Sky Survey. 

\noindent {\bf 1146+59}: This is the nearest and least radio-luminous galaxy in the sample. It has a double-lobed radio continuum structure with advance speeds of $(0.23\pm0.05)h_{100}^{-1}c$ \citep{tay98}, with \HI~detected in absorption against both jets \citep{pec98}. The separation speed between the jets is $(0.23\pm0.05)c$ with a separation of only $\sim2$~pc; if constant, this means the jets emerged only 34 years ago, making it the youngest known CSO. The host galaxy, NGC~3894, is classified as an E/S0 with an optical absolute magnitude of approximately $L^*$. HST observations with WFPC2 reveal two dust lanes nearly perpendicular to the radio axes, interpreted as the halves of a dusty torus at a high inclination angle \citep{per01}. The galaxy was observed with Chandra in early 2009 (Perlman et al., in prep).


\noindent {\bf 4C~+12.50}: The northern radio jet in this galaxy is much fainter and shorter than the southern counterpart, which extends out to $\sim150$~pc. There is a small amount of kpc-scale radio emission suggesting that the galaxy was active before the current burst of jet activity \citep{con91a}. The velocities of the northern jet components are $(1.0\pm0.3)h_{70}^{-1}c$ and $(1.2\pm0.2)h_{70}^{-1}c$, representing one of the only known examples of superluminal motion in CSOs; the velocity of the southern hotspot, however, is poorly constrained. Assuming a ``typical'' CSO advance speed of $0.3c$, \citet{lis03} calculate a kinematic age of 1700~yr; this is consistent with the dynamics of a precessing jet, requiring an age less than $10^5$~yr. \citet{dar02} also report a tentative detection of OH megamaser emission. Optical images of the host galaxy, IRAS~13451+1232, show irregular isophotes in a double nucleus, a tidal tail, and several companion galaxies indicating an advanced merger with little star formation \citep{gil86,sur98,sur99,sur00}. The low levels of star formation are puzzling, given the large amounts of molecular gas ($6.5\times10^{10}~M_\sun$) detected near the nucleus and its high IR luminosity \citep{mir89a,eva99}. High amounts of reddening and the presence of hard X-rays are interpreted as a quasar obscured behind a foreground dust screen by \citet{ode00}.  

\noindent {\bf OQ~208}: VLBI observations show a slightly asymmetric distribution between two radio lobes at a maximum separation of $\sim8$~pc, with no strong emission visible from the core \citep{sta97a}. The spectral turnover can be explained both by free-free and synchrotron self-absorption models, with lobe advance speeds of $0.33h_{70}^{-1}c$ \citep{kam00,xie05} and a kinematic age of 92~yr. The host galaxy, Mrk 668, shows the presence of both a tail and galactic companions in the optical \citep{sta93}, suggesting that the galaxy is dynamically disturbed. Complex Balmer line profiles classify the host galaxy as a broad line radio galaxy, or BLRG \citep{mar93}. \citet{gua04} detect flat X-ray emission from 2-10~keV with powerful $K_\alpha$ iron line emission, the first known example of a Compton-thick AGN in a BLRG. This is the only CSO in the sample with clear evidence of a broad-line region. 

\noindent {\bf PKS~1413+135}: This is the most radio-luminous and by far the most distant galaxy in the sample. At milliarcsecond scales, the galaxy has both a jet and counterjet with a strong core component, with hotspots extending out to $\sim139$~pc for the counterjet. The jets are likely oriented close to the line of sight; pressure balance arguments suggest an age of $10^3-10^4$~yr and velocities of $\sim0.05h_{100}^{-1}c$ \citep{per96,per02}. \citet{sto92} suggest that the radio component may be background to the optical counterpart, based on the absence of reprocessed radiation from near-IR dust continuum and emission lines. The host is a spiral galaxy with prominent dust lanes, a nuclear point source in the NIR, and a blue luminosity consistent with an Sbc classification \citep{per02}. \citet{wik97} detect several absorption lines in the spiral galaxy at $z=0.247$, consistent with a two-phase medium with dense molecular clumps embedded in a more diffuse component. \citet{per02} suggest that the absorbing screen consists of a giant molecular cloud (GMC) within the outer disk of the optical galaxy; this is consistent with OH absorption measured by \citet{dar04}.  


\noindent {\bf PKS~1718$-$649}: The radio emission has two lobes separated by $\sim2$~pc and no apparent emission from a core \citep{tin02}. The upper limit on the lobe separation speed is $0.08h_{100}^{-1}c$; \citet{tin97} adopt a kinematic age limit of $<10^5$~yr based on the jet energy and density advancing into an ISM with $n<3\times10^6$~cm$^{-3}$. The majority of the \HI~is distributed in an incomplete ring around the nucleus with a diameter of 37~kpc; kinematics of a weak \HI~envelope extending out to 180~kpc suggest a past merger with at least one gas-rich spiral galaxy \citep{ver95}. The total \HI~mass from VLBI is $3.1\times10^{10}M_\sun$. The host galaxy is NGC~6328, a LINER with photoionization as the dominant excitation mechanism \citep{fil85}. Optical images show a bright elliptical nuclear region and faint, extended spiral-like arms \citep{ver95}. A bar-like structure in the nuclear emission region imaged in H$\alpha$ and \nII~is coincident with the compact radio source \citep{kee91}.  

\noindent {\bf 1946+70}: The milliarcsecond radio structure shows an inverted core and two jets with multiple hotspots and ``S''-symmetry extending out to 25--35~pc \citep{tay97}. The expansion velocity between the hotspots is $0.024c$, corresponding to a kinematic age of $4000\pm1000$~years \citep{tay09}. Ejection of pairs of jet components appear to take place on timescales of $\sim10$~yr, with the most recent outburst observed in 1997. The peak \HI~absorption occurs near the core and extends outward somewhat into the jets, suggesting the presence of a foreground screen. \citet{pec99} propose a model in which the central jets are obscured by a torus of clumpy gas (both neutral and ionized) at 50--100~pc, with a warped molecular disk surrounding the AGN at $\sim1$~pc. HST observations of the host galaxy show mild isophotal twists and ellipticities, which may be indicative of a past merger \citep{per01}.  

\end{appendix}


\bibliography{cso,spitzer,general}

\end{document}